\documentclass[10pt,a4paper]{article}

\usepackage{flushend}
\usepackage{graphicx,color}
\usepackage[authoryear,round]{natbib}
\usepackage{amssymb,amsmath,stmaryrd}
\usepackage{mathrsfs}
\usepackage{multiequations}
\usepackage{subfigure}
\usepackage{amssymb,amsmath,stmaryrd}
\usepackage{mathrsfs}

\usepackage[colorinlistoftodos]{todonotes}
\usepackage{algorithm}
\usepackage{algpseudocode}
\usepackage[utf8]{inputenc}
\usepackage[french]{babel}
\usepackage[left=2cm,right=2cm,top=1.2cm,bottom=2cm]{geometry}

\newcommand{\mathsfbi}{\mbs}
\numberwithin{equation}{subsection}
\newcommand{\defi}{\stackrel{\triangle}{=}}

\newcommand{\id}{\mathbb{I}_d}
\newcommand{\transp}{^{\scriptscriptstyle T}}

\newcommand*{\be}{\begin{equation}}
\newcommand*{\ee}{\end{equation}}
\newcommand*{\bea}{\begin{eqnarray}}
\newcommand*{\eea}{\end{eqnarray}}
\newcommand*{\bal}{\begin{align}}
\newcommand*{\eal}{\end{align}}
\newcommand*{\bme}{\begin{multiequations}}
\newcommand*{\eme}{\end{multiequations}}

\providecommand\bcdot{\boldsymbol{\cdot}}

\newcommand\Rey{\mbox{\textit{Re}}}  

\newsavebox{\astrutbox}
\sbox{\astrutbox}{\rule[-5pt]{0pt}{20pt}}

\newcommand\eg{e.g.\ }

\def\squarebox#1{\hbox to #1{\hfill\vbox to #1{\vfill}}}

\newcommand{\defin}{\stackrel{\scriptscriptstyle\triangle}{=}}

\newcommand{\w}{\boldsymbol{w}}

\newcommand{\bphi}{\boldsymbol{\phi}}

\newcommand{\xx}{\boldsymbol{x}}

\newcommand{\XX}{\boldsymbol{X}}
\newcommand{\yy}{\boldsymbol{y}}

\newcommand{\nab}{\boldsymbol{\nabla}}

\newcommand{\Exp}{\mathbb{E}}

\newcommand{\dif}{{\mathrm{d}}}
\newcommand{\mbs}[1]{\ensuremath{\boldsymbol{#1}}}

\begin{document}

\title{Stochastic modelling  and diffusion modes for POD  models  and small-scale flow analysis}

\author{
Valentin Resseguier$^{1,2}$\thanks{Email address for correspondence: valentin.resseguier@inria.fr}, Etienne M\'emin$^1$,\\
Dominique Heitz$^{3,1}$ and Bertrand Chapron$^2$
\\
	$^1$Fluminance, Inria/Irstea/IRMAR,\\
	Campus de Beaulieu, 35042 Rennes, France\\
	$^2$LOPS, Ifremer, Pointe du Diable, 29280 Plouzan\'{e}, France\\
	$^3$Irstea, UR OPAALE, F-35044 Rennes, France\\
}

\maketitle   

\fontsize{9}{11}\selectfont

\section*{ABSTRACT}

We present here a new stochastic modelling in the constitution of fluid flow reduced-order models. This framework introduces a spatially inhomogeneous random field to represent the unresolved small-scale velocity component. Such a decomposition of the velocity in terms of a smooth large-scale velocity component and a rough, highly oscillating component gives rise, without any supplementary assumption, to a large-scale flow dynamics that includes a modified advection term together with an inhomogeneous diffusion term. Both of those terms, related respectively to {\em turbophoresis} and {\em mixing} effects, depend on the  variance of the unresolved small-scale velocity component. They bring an explicit subgrid term to the reduced system which enable us to take into account the action of the truncated modes. Besides, a decomposition of the variance tensor in terms of {\em diffusion modes} provides a meaningful statistical representation of the stationary or nonstationary structuration of the small-scale velocity and of its action on the resolved modes. This supplies a useful tool for turbulent fluid flow data analysis. We apply this methodology to circular cylinder wake flow at Reynolds numbers $Re=100$ and $Re=3900$, respectively. The finite dimensional models of the wake flows reveal the energy and the anisotropy distributions of the small-scale diffusion modes. These distributions identify critical regions where corrective advection effects as well as structured energy dissipation effects take place. In providing rigorously derived subgrid terms, the proposed approach yields accurate and robust temporal reconstruction of the low-dimensional models.

\section{Introduction}\label{sec:intro}

   Surrogate empirical models of flow dynamics with a reduced set of degrees of freedom  are widely used in fluid mechanics for control applications or physical analysis \citep{Noack-Book10}. Within such modelling a few numbers of modes extracted  from experimental or numerical measurements are used to represent the main dynamical behaviour of a flow. The modes in themselves may help in unveiling recurrent dynamical patterns. Spectral approaches are quite natural for this purpose. Fourier representation has been used for a long time to characterize hydrodynamic instabilities. Proper Orthogonal Decomposition (POD) and the spectral representation of the velocity auto-correlation  matrix are used to extract descriptive empirical spatial or temporal basis of the flow \citep{Aubry88,Holmes96,Sirovich87}. More recently the Dynamic Modes Decomposition (DMD) relying on the eigenvectors of the Koopman operator \citep{Koopman31} and Takens's delay embedding theorem \citep{Takens81} have been proposed to represent,  from the evolution of observations,  the principal modes of the  dynamical system's attractor \citep{Mezic05,Rowley09,Schmid10}. Combination of both representations can also be used to provide a suitable energy spectrum representation \citep{Cammilleri-TCFD13}. In all those modal representations the construction of the reduced-order dynamics requires a truncation operation in which the most ``influential''  modes -- with respect to a given criterion -- are kept  to describe the flow. In general, the action of the discarded modes must be  modeled to get accurate and stable dynamical systems.  The effect of those  neglected processes encompasses dissipation effects  but is also responsible for some energy redistribution and backscattering \citep{Piomelli-91}.

In most of the flow low-order dynamics, the unresolved small-scale processes are represented on the basis of an eddy viscosity assumption \citep{Boussinesq77}. This takes the form of  a damping term in the reduced-order dynamical system. In Galerkin POD  reduced models, this extra dissipation,  which is added to the linear molecular diffusion, is modeled by a constant coefficient \citep{Aubry88} or through a modal constant vector \citep{Cazemier98, Rempfer94}. Recently, nonlinear functions have been proposed for a bluff body wake flow \citep{osth2014need}. Although those models have demonstrated their efficiency in numerous situations, the estimation of the associated parameters and/or the choice of the nonlinear dependency between the eddy-viscosity coefficients and the modal coefficients constitute a sensible issue. Furthermore, from a physical interpretation point of view, the action of the small-scale velocity component is interpreted only with regard to a homogeneous stationary dissipation effect. Neither preferential local direction of diffusion related to the flow physics, nor energy redistribution action by the small scales are considered.

  Robust techniques based on optimal control strategies have also been proposed for building reduced dynamical models from noisy data \citep{Dadamo-JOT07,Artana-JCP12,Cordier_etal_2013,Semaan16} and incomplete knowledge of the actual flow dynamics (i.e. unknown initial condition,  partially known forcing terms, etc.). 
  Calibration techniques built from very close paradigms have also been proposed to tune appropriately dynamics parameters \citep{couplet2005calibrated,buffoni2006low,perret2006}.  Those techniques accurately estimate low-order dynamical systems in the temporal windows on which the data are available. 
These methods unfortunately  present some limitations for predicting new states of the system. Furthermore, the  physical interpretation of the unresolved velocity component remains difficult since its contribution is distributed in an unknown manner over all the coefficients of the dynamical system and on the error function when weak dynamical constraint is considered \citep{Artana-JCP12}. 
 
 In this work, to take into account the unresolved modes in the surrogate dynamic model, we will rely on a recently proposed stochastic framework \citep{memin2014fluid,resseguier2016geo1}. In this context, an advection of the large-scale component due to the action of the unresolved random component emerges naturally, together with an inhomogeneous nonstationary diffusion. This will lead us to consider corrective advection and diffusion terms driven by the turbulence inhomogeneity whose local effects can now be physically interpreted.
 
 After presenting the stochastic model in \S \ref{Dynamics stochastic modelling}, we describe  the derivation of the associated POD reduced-order model in \S \ref{Stochastic reduced-order model with differentiable velocity}. Furthermore, we propose a method to estimate the additional components of the dimensional reduced system from the residual velocity. Then, the data benchmarks are detailed in \S \ref{sec:flow configuration}. From the estimated additional components, we analyse  the influence of the residual velocity on the large-scale flow and reconstruct the temporal modes of the reduced-order models in \S \ref{Numerical results}.

\section{Dynamics stochastic modelling}
\label{Dynamics stochastic modelling}

The proposed stochastic  principle relies on a Lagrangian random description of the flow velocity:
\begin{equation}
\frac{\dif \XX_t}{\dif t} = \w(\XX_t,t) + \dot{ \mbs{\eta}}(\XX_t,t).
\label{particle_dX}
\end{equation}
The first right-hand term, $\w$, stands for the large-scale velocity component.  
It is a smooth component  along time. For turbulent flows, it is associated with a much larger time-scale than the unresolved small-scale velocity component.
This latter, $\dot{ \mbs{\eta}}=\dif \mbs{\eta} /\dif t$, is associated with fast modes that are rapidly decorrelating at the resolved time scale. Based on this observation, we will assume that such a component can be  ideally represented through  a spatially smooth incompressible (divergence-free) Gaussian random field uncorrelated in time.  This  (possibly inhomogeneous) random field is formally built from an infinite-dimensional Brownian motion. It is associated with a covariance tensor denoted:
\begin{equation}
 \mathsfbi Q_{ij}(\xx,\yy,t,t') =\Exp (\dif \eta_i(\xx,t) \dif \eta_j(\yy,t'))= \mathsfbi c_{ij}(\xx,\yy,t)  \delta(t-t') \dif t.
\end{equation}
In the following, the diagonal of the covariance tensor, which plays a central role in our setting, will be denoted  as: $\mathsfbi  a (\xx) \defin \mathsfbi  c(\xx,\xx,t)$. This  tensor, that may depend on time, will be referred to as the small-scale variance tensor. It is a symmetric positive definite matrix at all spatial points, $\xx$ (excluding degenerate cases) with dimension in $m^2.s^{-1}$. It thus corresponds  to an eddy viscosity term.

This stochastic formulation is related in spirit to the Lagrangian stochastic models based on Langevin equations  that have been intensively used for turbulent dispersion \citep{Sawford86} or in probability density function (PDF) modelling  of turbulent flows \citep{Haworth86,Pope94, Pope00}. However,  our interest here focuses  on the associated large-scale Eulerian representations of the flow dynamics.  This Eulerian description of the resolved velocity component is obtained through a formulation of the Reynolds transport theorem adapted to such a stochastic flow.

\subsection{Stochastic conservation equations}
Considering the flow decomposition (\ref{particle_dX}), the rate of change of a scalar quantity (in the absence of random forcing) within a material volume is given by the following expression  \citep{memin2014fluid,resseguier2016geo1}:
\begin{eqnarray}
\label{th_transport}
\frac{\dif}{\dif t}
 \int_{V (t)} q
\;  \dif \xx =
   \int_{V (t)}  \left ( 
\frac{\partial q}{\partial t}
+  \nab\bcdot ( q \w^*)
- \nab\bcdot \left( \frac{1}{2} \mathsfbi a\nab q \right)
 +  
 \dot{ \mbs \eta}
 \bcdot 
 \nab q
  \right)  \dif \xx ,
  \end{eqnarray}
 where the effective advection velocity is given as:
\begin{equation}
\w^* \defi \w - \frac{1}{2} \bigl(\nab\bcdot \mathsfbi a \bigr)\transp.
\label{ef-adv}
\end{equation} 
Equation (\ref{th_transport}) provides a stochastic representation of the so-called Reynolds transport theorem. It is important to outline that at a given grid point,  $q$  is a random value which depends among other things on the Brownian component of the particles flowing through that point. The second term corresponds to the large-scale advection by an effective drift, $\w^*$, that includes a contribution related to the divergence of the small-scale velocity variance tensor (\ref{ef-adv}). The third term is a diffusion expressing the mixing effect exerted by the small-scale velocity component.  The final term corresponds to  the scalar advection by the small-scale  velocity field.   
 From this expression a conservation of an extensive property, $\int_{V (t)} q$, such as mass or internal energy (neglecting diabatic and compressive effects) reads immediately as the following intensive property evolution equation
\begin{equation}
\label{SST}
\frac{\partial q}{\partial t}
+  \nab\bcdot\left ( q \w^*\right)
+
 \dot{ \mbs \eta}
 \bcdot 
 \nab q
= \nab\bcdot  \left ( \frac{1}{2} \mathsfbi a \nab q  \right) 
 .
\end{equation}
As the right-hand term is a smooth temporal component, we observe immediately that the Brownian terms associated, on the one hand, to the scalar temporal variation and, on the other hand, to the small-scale advection necessarily compensate each other. A fluid with a constant density $\rho$, naturally requires  a divergence-free constraint on the effective advection:
\begin{eqnarray}
\label{eq_incomp_sigma}
0 &=& \nab\bcdot
 \dot{ \mbs \eta}
 , \\
\label{eq_incomp_w*}
0 &=&  \nab \bcdot \w^* = \nab\bcdot \left ( \w  - \frac{1}{2} ( \nab\bcdot \mathsfbi a)\right) .
\end{eqnarray}
  This is the case we are dealing with  in this study. The two constraints (\ref{eq_incomp_sigma}--\ref{eq_incomp_w*}) correspond to the incompressibility conditions associated with the stochastic representation. For isochoric flows with variable density as in geophysical fluid dynamics, interested readers can refer to \cite{resseguier2016geo1,resseguier2016geo2,resseguier2016geo3}.
  

\subsection{Navier-Stokes equations associated with a stochastic representation of the small-scales}
\label{1stNavierStokesstochasticmodel}
As in Newton second law, a dynamical balance between the temporal differentiation of the stochastic momentum, $\rho \dif \XX_t$, and the action of the forces is assumed. Applying the stochastic representation of the Reynolds transport theorem (\ref{th_transport}) leads to the following Navier-Stokes equations \citep{memin2014fluid}:
\begin{equation}
\label{NSvf}
\frac {\partial \w} {\partial t} + (\w^* \bcdot \nab ) \w   
=  - \frac 1 \rho \nab p 
+ \sum_{i,j=1}^d 
\frac{\partial }{\partial x_i}
 \left (\frac{1}{2} 
 \mathsfbi a_{ij} 
  \frac{\partial  \w}{\partial x_j}  \right ) + \nu \triangle \w.
\end{equation}
This equation corresponds to the large-scale momentum equation. This expresssion differs from the classical Reynolds decomposition  formulation mainly by the introduction of both a large-scale dissipation term and a correction term in the large-scale advection.  The dissipative term plays a role that is similar to the eddy viscosity models which are introduced in classical large-scale representations \citep{Bardina80,Lilly92,Smagorinsky63} or to spectral vanishing  viscosity \citep{Karamanos00,Pasquetti06,Tadmor89}. It is also akin to numerical regularization models considered in implicit models \citep{Aspden08,Boris92,Lamballais11}. The small-scale stochastic representation principle  is nevertheless  more general as it does not rely on {\em a priori} fixed shapes of the subgrid tensor (\eg Boussinesq assumption) nor does it  presuppose a given numerical scheme (\eg implicit models). The subgrid term takes a general diffusion form whose matrix coefficients are given by the small-scale variance tensor. The diffusion principal directions are thus aligned with this tensor principal directions. 

The advection correction term is much less intuitive. It is related here to an advection bias due to the inhomogeneity of the small-scale variance tensor. This corresponds to the eddy-induced velocity introduced for tracer mean transport in oceanic or atmospheric circulation models \citep{Andrews76,Gent95}  and more generally to the {\em turbophoresis} phenomenon associated with small-scale inhomogeneity, which drives inertial particles toward the regions of lower diffusivity \citep{Brooke92,Caporaloni75,Reeks83,Sehmel70}. Qualitatively, this drift correction can be understood as follows. Fluid parcels with higher turbulent kinetic energy (TKE) move faster. It ensues that at large scales, areas associated with maximum TKE spread whereas areas associated with minimum TKE shrink. Hence, a large-scale drift oriented toward these maxima/minimum emerges. This orientation suggests an anticorrelation with the TKE gradient. Since the turbulent velocity variations are multidimensional, they are better described by the variance tensor. The drift correction is consequently proportional to the opposite of the variance tensor divergence. For homogeneous turbulence, the small-scale variance tensor is constant and this corrective advection does not come into play. 
It can be noted that this advection correction is of the same form as that proposed in \citet{Caporaloni75,Macinnes92,Reeks83}. 

The small-scale random field can be freely defined and be in a shape that goes from isotropic stationary models up to inhomogeneous non-stationary random fields. However, in the inhomogeneous case (such as the Smagorinsky model)  the  advection correction term comes into play. A stochastic representation  of the unresolved scales thus differs significantly from classical large-scale modelling. It relies on less strict assumptions,  which enable us to cope naturally with inhomogeneous anisotropic turbulence.

This stochastic representation relies on a  scale gap assumption, which is coherent with deterministic justifications of the eddy viscosity \citep{kraichnan1987eddy}. The stochastic transport expression (\ref{th_transport}) and the momentum equation (\ref{NSvf}) provide the foundations of a physically relevant large-scale fluid dynamics formulation. It opens a new paradigm for large-scale modelling adapted to turbulence inhomogeneity in involving a general subgrid diffusion together with a small-scale drift correction. In the next section, we will rely on this model for the construction of reduced-order dynamical systems. 


\section{Reduced-order models}
\label{Stochastic reduced-order model with differentiable velocity}
 Dimensional reduction techniques enable the constitution of simplified lower dimensional representations of partial differential equations (PDE). They are usually specified from a Galerkin projection onto data-based dedicated basis. The proper orthogonal decomposition, also called Empirical Orthogonal Functions (EOF) in geophysics, is one of those methods for turbulent flows. In \S \ref{classical POD} the POD model reduction is briefly presented. Then, in \S \ref{ssec:romstoc} we introduce the derivation of the reduced-order model from the stochastic representation principle described in \S \ref{Dynamics stochastic modelling}. In \S \ref{Choice of the time step}, different characteristic time scales are introduced for the different modes, leading to the concept of modal characteristic time steps. Finally, in \S \ref{Estimation of the tensor $a$} a precise specification of the small-scale variance tensor is proposed with two different estimation methods.

\subsection{POD model reduction}
\label{classical POD}
POD reduced-order models rely on the linear decomposition of the velocity $\w$  on a reduced number  of orthogonal spatial modes (\cite{Holmes96}):
\begin{eqnarray}
\label{approx_POD}
\w(\xx,t) \approx  b^i(t) \mbs \phi_i (\xx),
\end{eqnarray}
where we used Einstein summation convention. Unless stated otherwise, this convention is adopted in this paper.

The number of modes, $n$,  is assumed to be much lower than the state space dimension. The functions $\left (\mbs \phi_i(\xx) \right )_{1 \leqslant  i \leqslant N}$ encoding the spatial flow variations are referred to as \textit{topos} and are computed from a Karunen-Loeve decomposition on a series of $N+1$ available velocity snapshots. The \textit{topos} are sorted by decreasing order of the snapshots' empirical covariance eigenvalues: $ \lambda_1 > ...  >  \lambda_N  $. The $\left (b_i(t) \right )_{1 \leqslant  i \leqslant N}$ denote the temporal modes; they are called \textit{chronos}. The {\em chronos} are the eigenvectors of the spatially averaged temporal correlation matrix, whereas the {\em topos} constitute the eigenvectors of the temporally averaged spatial correlation matrix. They are both computed from the snapshots' covariance. Function $\phi_0$ corresponds to  the time average velocity and $b_0 \defi \lambda_0 \defi 1$. We also denote by $T$ the time between the first and the last snapshot. The Navier-Stokes equations can be written in  the general following form:
\begin{eqnarray}
\label{compact_NS}
 \frac {\partial \w} {\partial t} = \mbs I + \mbs L(\w)+ \mbs C(\w,\w),
 \end{eqnarray}
where $\mbs L$ and $\mbs C$ stand respectively for  linear and bilinear differential operators. The first term, $\mbs I$, collates the  pressure and the external forces such as gravity. The second one, $\mbs L$, includes the molecular friction term and possibly the Coriolis force. The last one, $\mbs C$, encodes the nonlinear advection term.
Projecting this PDE on each \textit{topos} (with the $\mathcal{L}^2$ scalar product noted $\langle\bcdot,\bcdot\rangle$):
\begin{equation}
\bigl\langle \partial_t w, \phi_i\bigr\rangle =
\bigl\langle \mbs I, \phi_i\bigr\rangle + \bigl\langle \mbs L(\w), \phi_i\bigr\rangle + 
\bigl\langle \mbs C(\w,\w), \phi_i\bigr\rangle,
\end{equation}
leads to a system of ordinary differential equations for the \textit{chronos}: 
\begin{equation}
\label{classic POD}
\frac {\dif  b_i} {\dif t}
 = \underbrace{\left ( \int_{\Omega} \mbs {\phi}_i  \bcdot  \mbs I   \right ) }_{\defi i_i}
+ 
 \underbrace{ \left ( \int_{\Omega}  \mbs {\phi}_i  \bcdot \mbs L(\mbs {\phi}^p)  \right ) }_{\defi l_{i}^p} b_p  
 +
  \underbrace{\left ( \int_{\Omega}  \mbs {\phi}_i  \bcdot \mbs C(\mbs {\phi}^p,\mbs {\phi}^q)  \right ) }_{\defi  c_{i}^{pq}} b_p b_q .
\end{equation}

Due to nonlinearity, the temporal modes strongly interact with one another. In particular, even though the original model (with $n=N$) is computationally stable for moderate Reynolds number, a strongly reduced model ($n\ll N$)  appears unstable in general. A frequency shift is also often observed. Those artefacts  are extensively documented  in the literature  \citep{Artana-JCP12, Aubry88,Rempfer94,osth2014need,protas2015optimal}. 
The introduction of a damping  eddy viscosity term  to  mimic the truncated modes' dissipation leads to a modified linear term in \eqref{classic POD}. Unfortunately, as this term is built on empirical grounds its precise form is difficult to justify. Furthermore, its parametrization has to be tuned for each simulation to achieve good results. When large wake domains are considered the influence of the pressure term (and of the boundaries) is in general negligible \citep{Deane91,Ma03, Noack05}. We will also rely on this assumption, although several authors have shown that neglecting the pressure term was a source of uncertainty regarding an accurate representation of  the flow dynamics \citep{Kalb07,Noack05}. To take into account the effect of the outflow boundary, corrective terms are introduced by some authors through modifications of the linear \citep{Galetti07} or quadratic terms \citep{Noack05}.
\subsection{Reduced-order modelling associated with the stochastic representation}\label{ssec:romstoc}
To overcome the difficulties previously evoked, we propose to derive the reduced-order model from the stochastic representation principle described previously. 
To account for the effect due to the modal truncation, we will assume that the whole  field $\mbs u = \w + \dot{\mbs {\eta}}$ can be decomposed in such a way that the large-scale component lives on the subspace endowed with the reduced POD basis $\w= \sum_{i=0}^{n} b_i \mbs{\phi}_i $ while realisations of the small-scale component  belong to the orthogonal complement subspace  $\dot{\mbs \eta} =\sum_{i=n+1}^{N} b_i \mbs{\phi}_i $. Since $\nab \bcdot \mbs u =0$, for all $i$, $\nab\bcdot \bphi_i=0$ and, then, $\nab \bcdot \w=0$. The dynamics of the large-scale component, $\w$, is given by  the incompressible Navier-Stokes equations (\ref{NSvf}).
Projecting this equation onto the {\it topos}  $\mbs{\phi}_i$ now leads
for $i>0$ to:
\begin{eqnarray}
\label{eq_evol_b}
\frac {\dif  b_i} {\dif t} 
&=& i_i + \left (  l_{  i}^p+ \breve{ f}_{ i}^p(\mathsfbi a) \right ) b_p  +  c_{ i}^{pq} b_p b_q \ , \\
\label{f definition}
\text{with }
 \breve{ f}_{ i}^p(\mathsfbi a)
& \defi &
  \int_{\Omega}  
{\phi}^{(k)}_i
      \ \Biggl( 
  \underbrace{
-  \frac 12 (\nab \bcdot \mathsfbi a) \nab \phi^p_{(k)} 
}_{\text{Advection}}
+   \underbrace{
 \nab\bcdot \left ( \frac{1}{2} \mathsfbi a \nab \phi^p_{(k)}  \right )
 }_{\text{Diffusion}}
 \Biggr ) ,
\end{eqnarray}
where $\phi^{(k)}_i$ and $\phi^p_{(k)} $ stand for the $k$-th coordinate of the $i$-th and of the $p$-th \textit{topos} respectively.

The additional term $\breve{\mbs f}(\mathsfbi a)$ corresponds to the projection on the \textit{topos} of the effective advection and the diffusion brought by the stochastic representation of the small scales. We note that this function is linear and is the only function that depends on the variance tensor $\mathsfbi a$.
This system now includes  a natural small-scales dissipation mechanism, through the diffusion term. But it also corrects the frequency shift through the additional advective term 
ensuing from the small-scale inhomogeneity.

To fully define this system, we need to specify the  small-scale variance tensor $\mathsfbi a$. This issue is developed in subsection \ref{Estimation of the tensor $a$}. But before that, we will further elaborate on the choice of the characteristic times related to the modal truncation.

\subsection{Time scale characterisation}
\label{Choice of the time step}
Very efficient  flow simulations are obtained by reducing as much as possible the number of modes of the associated surrogate model. An even higher efficiency can be obtained by increasing the evolution time step. 
Moreover, a larger sampling time step of the snapshots also reduces the computational cost of the reduced system computation.
This  time step can be naturally chosen as a single constant for the whole system. However, as we shall see, different characteristic time scales can be fixed for the different modes, leading to the concept of modal characteristic time steps.
 
\subsubsection{Single time step}
As long as the resolved modes, representing $\w$, are smooth  with respect to time, the assumption pertaining to our reduced model construction is valid. The time-step must thus be fixed as the largest value that guarantees that all the {\it chronos}  remain smooth.   The characteristic time scale  associated with the fastest resolved mode (which is often the least energetic mode) is a good target for that purpose. This time scale is associated with the highest frequency of the {\em chronos} Fourier modes.  Quantitatively, the Shannon-Nyquist theorem provides us with a natural upper bound to fix its value. This theorem states that a function can be sampled, without loss of information, if the sampling frequency is at least twice as large as the largest frequency of the original function. Otherwise, the sampled function undergoes an aliasing artifact characterized by a folding of the Fourier spectrum and a loss of regularity. We will thus assume that the  required regularity condition is fulfilled if the modes are not affected by aliasing phenomena. A sufficient condition thus reads:
\begin{equation}
\label{eq time step}
\frac{1}{\Delta t} \geqslant  2 \max_{i \leqslant n}  \left ( f_{max} \left ( b_i \right ) \right ).
\end{equation}
where $f_{max} \left ( b_i \right )$ is the maximum frequency of the $i$-th mode. Aliasing takes place in the unresolved temporal modes, which are associated with smaller time scales
, and live on the \textit{chronos} complement space. 
However, the stochastic representation  is precisely built from a decorrelation assumption of the small-scale unresolved part of the velocity. This characteristic time scale of the resolved modes may be thus also seen as a sampling time at which the unresolved modes appear uncorrelated. So, a strong subsampling of those components strengthens further the decorrelation property of the unresolved modes.

\subsubsection{Modal characteristic times}
The previous model can be enriched by considering that the resolved \textit{chronos} are associated with different time scales.  So, we introduce a new criterion that reads for
  each \textit{chronos}, $b_i$:
\begin{equation}
\label{eq modal time step}
\frac{1}{\Delta t_i} \geqslant   2   f_{max} \left ( b_i \right ) .
\end{equation}
A modal variance tensor field for each \textit{chronos} 
immediately follows:
\bea
\mathsfbi a^{(i)}(\xx) 
\defi
\Delta t_i
\Exp \left \{ \dot{\mbs \eta} \dot{\mbs \eta}\transp \right \}(\xx)
=
 \frac{\Delta t_i}{\Delta t} 
 \tilde {\mathsfbi a} (\xx)
 \text{ with }
  \tilde{\mathsfbi a}(\xx)
  =
\Delta t \ 
\Exp \left \{ \dot{\mbs \eta} \dot{\mbs \eta}\transp \right \}(\xx).
\eea
 The modal variance tensor $\mathsfbi a_i$ corresponds to the small-scale velocity variance 
 over a period given by 
 the time-scale of the \textit{Chronos} $b_i$ ({i.e.} it corresponds to an eddy  viscosity associated with the neglected modes expressed with respect to the characteristic time associated with $b_i$). The {\em chronos} evolution equation \eqref{eq_evol_b} becomes:
\begin{eqnarray}
\label{eq_evol_b modal dt}
\frac{\dif b_i}{\dif t}  
&=&
 i_i 
 +
  \left (  l_{  i}^p+  \frac{\Delta t_i}{\Delta t}  \breve{ f}^p_i(  \tilde{\mathsfbi a} ) \right ) b_p
 +
 c_i^{pq} b_p b_q 
 \text{\hspace{1cm} (no sum on }i)
 .
\end{eqnarray}
In practice, the common characteristic time $\Delta t$  is set up by formula \eqref{eq time step}. 
Even though it does not depend on the index $i$ of the \textit{chronos}, this time is a function of the number $n$ of resolved modes.
The estimation procedures involved in the construction of the reduced system -- detailed in the following -- use this characteristic time as a sampling time step for the velocity snapshots. Let us note that to reconstruct the \textit{chronos} from that reduced system, the simulation time step is in general fixed to smaller values than this characteristic time; it is indeed ruled by a CFL condition in order to ensure the convergence of the temporal discretization scheme.

\subsection{Estimation of the small-scale variance tensor}
\label{Estimation of the tensor $a$}

The full definitions of the reduced-order 
models \eqref{eq_evol_b} and \eqref{eq_evol_b modal dt} require
 a precise specification of the small-scale variance tensor. We compare here two different estimation methods for this tensor.  A first method will rely on a stationarity assumption while a second technique will allow us to define a  non-stationary tensor. 
 To avoid misunderstanding, note that the time steps $\Delta t$ and $\Delta t_i$ -- previously described -- are \textit{a priori} not related to the possible time variations of the variance tensor. Indeed, the characteristic time steps $\Delta t$ and $\Delta t_i$ are by definition associated with the fastest time variations of the resolved \textit{chronos}. In contrast, the possible time variations of the variance tensor is defined as the time evolution of the variance of the unresolved modes.

\subsubsection{Stationary small-scale variance tensor}
This case corresponds to the model developed in \cite{memin2014fluid}.
The small-scale velocity variance, $\mathsfbi a / \Delta t$ can be computed  through a temporal  averaging of the residual velocity second moment $(\mbs v-\w)(\mbs v-\w)\transp(t_i)$ at all spatial locations. This simple scheme thus provides  a  representation of a spatially varying  stationary variance tensor. 
The computational cost of this estimator linearly depends on the number of snapshots and is thus inversely proportional to the sampling characteristic time step $\Delta t$ of the unresolved modes.

\subsubsection{Small-scale variance tensor in the chronos subspace}
A stationary model has obvious limitations in terms of  turbulence intermittency modelling.
A time dependent  variance tensor is nevertheless more complex to estimate as in this case only a single  realization of the small-scale velocity trajectory, $(\mbs v(\xx,t)- \w(\xx,t))$, is available.
However considering a temporal basis it is possible  to estimate, at a fixed point, the matrix coefficients, $\mathsfbi z_i(\xx)$, of the tensor, $\mathsfbi a(\xx,t)$ \citep{genon1992non}. We term those coefficients the {\em diffusion modes}, as they correspond to a modal decomposition of the principal diffusion directions. With  the \textit{chronos} reduced basis we get:
\begin{equation}
\label{eq a span chronos}
\tilde {\mathsfbi a}(\xx,t) =  b^j(t) \mathsfbi z_j(\xx).
\end{equation}
Note that even though the residual velocity, $(\mbs v - \w)$, lives in the subspace which is orthogonal to the \textit{chronos} reduced basis, its one-point one-time covariance -- and hence the variance tensor -- do not. So, it seems natural to introduce the decomposition \eqref{eq a span chronos}.
Using the orthogonality of the \textit{Chronos}, $\int_0^T b_k b_l\dif t = \delta_{kl} \lambda_k T$ (no sum on $k$), leads without sum on $j$ to:
\begin{multline}
\label{estim formula of z_i}
\mathsfbi z _j(\xx) = \int_0^T \frac{b_j(t)}{T \lambda_j} \tilde {\mathsfbi a}(\xx,t) \dif t
 \approx 
  \frac{ \Delta t }{N+1} \sum_{k =0}^{N} 
  \frac{b_j (t_k)}{\lambda_j} (\mbs v-\w)(\xx,t_k) ((\mbs v-\w)(\xx,t_k) )\transp  
 ,
\end{multline}
where $N = \frac{T}{\Delta t }$. Again, the computational costs of these estimators are inversely proportional to the sampling time step $\Delta t$.
The formulas \eqref{eq a span chronos} and \eqref{estim formula of z_i} -- rigorously supported by stochastic calculus theory \citep{genon1992non} -- can be heuristically understood as a time smoothing of the square residua, $(\mbs v-\w)(\mbs v-\w)\transp$. Indeed, by the projections of this  square residua onto the large-scale \textit{chronos} we only keep the large-scale patterns of the  square residua.
It can be noticed that keeping only the zero-diffusion mode and cancelling  the others:  $z_i=0,\; \forall  i \geqslant 1$, brings us back  to the stationary variance tensor model. The non-zero modes introduce a non-stationary variance. Yet, it is important to outline that  the reduced-order model (\ref{eq_evol_b}) remains a quadratic autonomous system. As a matter of fact from (\ref{eq a span chronos}), we get the following system:
\begin{equation}
\label{avant derniere eq dbi}
\frac{\dif b_i}{\dif t}  =
 i_i 
 +
   l_i^p b_p 
 +
 \left ( c_i^{pq} +    f_i^{pq}  \right ) b_p b_q  ,
\text{ where } f_i^{pq} \defi \frac{\Delta t_i}{\Delta t} \breve{ f}_i^q \left( {\mathsfbi z}^p \right)
\text{\hspace{0.5cm}(no sum on $i$)}
.
\end{equation}

Finally, we can combine the parametrization based on single or modal characteristic times with the stationary or with the non-stationary variance tensor model. As such, we obtain four distinct methods. Even with their large adaptabilities, each of these methods only requires simple estimation algorithms and yields an autonomous quadratic reduced order model.


\section{Flow configuration and numerical simulations}\label{sec:flow configuration}

To evaluate the pertinence of the modelling developed in the previous section for the specification of a low order dynamical system and to analyse the contribution of the small-scale component, we consider two-dimensional and three-dimensional incompressible flows past a circular cylinder at Reynolds number $Re=100$ and $Re=3900$ respectively. 
For the Reynolds $Re=100$, we  performed direct numerical simulations (DNS) using Incompact3d, a high-order flow solver, based on the discretization of the incompressible Navier--Stokes equations with finite-difference sixth-order schemes on a cartesian mesh \citep{Laizet09}. A second-order Adams-Bashforth scheme was used for the time advancement. The incompressibility condition is treated with a fractional step method based on the resolution of a Poisson equation in spectral space, allowing here for the velocity field the use of periodic boundary conditions in the two lateral directions $y$ and $z$. A constant flow is imposed at the inlet of the computational domain and a simple convection equation is solved at the exit. Using the concept of the modified wavenumber, the divergence-free condition is ensured up to machine accuracy. The pressure field is staggered from the velocity field by half a mesh to avoid spurious oscillations.  
The modelling of circular cylinder of diameter $D$ inside the computational domain was performed here with a simple Immersed Boundary Method (IBM). It is based on a direct forcing to ensure zero velocities boundary condition at the wall and inside the solid body. 
We performed the DNS at Reynolds number $Re=100$ on a domain extending over  $20D \times 20D \times 0.5 D$ with 
$241 \times 145 \times 8$
 points in the streamwise, perpendicular and spanwise directions, respectively. 
This reduced spanwise length corresponds to the minimum domain size usable with Incompact3d and led to a three dimensional wake flow simulation with a very short periodicity in the spanwise direction. 
 We highlight the fact that the third dimension is here only for a practical numerical reason.
 Incompact3d cannot be used without this spanwise direction. However, no $3$-dimensional structure are present here.
This low Reynolds choice was made to reduce the computational cost and to simulate a longer time series, necessary for the POD analysis. 
 $N=10,000$ snapshots are saved to observe $100$ vortex shedding cycles.

In addition, a large-eddy numerical simulation (LES) was performed with Incompact3d.
This code solved incompressible Navier--Stokes equations on a  grid stretched along the $y$ direction in nonstaggered configuration. It uses the customized IBM technique of \citet{gautier2014} to avoid discontinuities on the velocity field, leading to the creation of spurious oscillations when high-order centred schemes are used. 
Except the grid stretching and the  resolution, the simulation configuration is similar to that described in \citet{parnaudeau2008experimental}. The subgrid-scale model proposed by \citet{Smagorinsky63}  was combined with a fixed filter length which is estimated as the cubic root of the mesh volume. The subgrid parametrization is a classic Smagorinsky subgrid model with the constant $C_s=0.1$ as suggested by \cite{ouvrard2010classical}. To provide long time integration data, this LES was carried out with the low spatial resolution configuration used by \citet{parnaudeau2008experimental}. The LES  was computed on a domain size of $20D \times 20D \times \pi D$ with  $481 \times 481 \times 48$ points in the streamwise, perpendicular and spanwise directions, respectively. We extracted from this simulation $1,460$ equidistant snapshots over $73$ vortex shedding cycles.

\begin{table}
  \begin{center}
\def~{\hphantom{0}}
  \begin{tabular}{lcccccc}
      Case  & $Re$   &   $(L_x \times L_y \times L_z)/D$ & $n_x \times n_y \times n_z$ & Stretching & Snapshots & Sheddings \\[5pt]
       DNS   & 100 & $20 \times 12 \times 0.5 $ & $241 \times 145 \times 8$ & None & 10,000 & 100\\
       LES   & 3900 & $20 \times 20 \times \pi$ & $481 \times 481 \times 48$ & Along $y$ &  1,460 &  73\\
  \end{tabular}

  \caption{Summary of simulations and of extracted data.}
  \label{tab:cases}
  \end{center}
\end{table}

\begin{figure}
\centering
\includegraphics[width=5in]{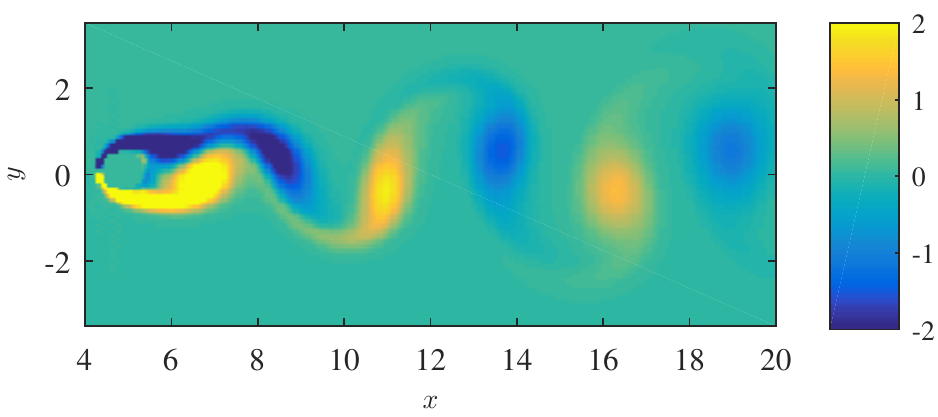}\\
\includegraphics[width=5in]{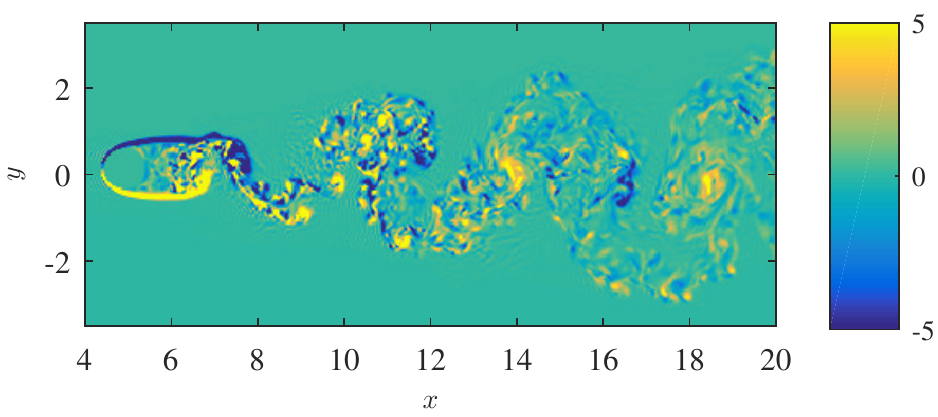}\\
\caption{
 Spanwise vorticity component in a circular cylinder wake flow at Reynolds number, from top to bottom,  $Re=100$ (DNS) and $Re=3900$ (LES) respectively.}
\label{fig_vorticity100}
\end{figure}

\begin{figure}
\centering
\includegraphics[width=5.5in]{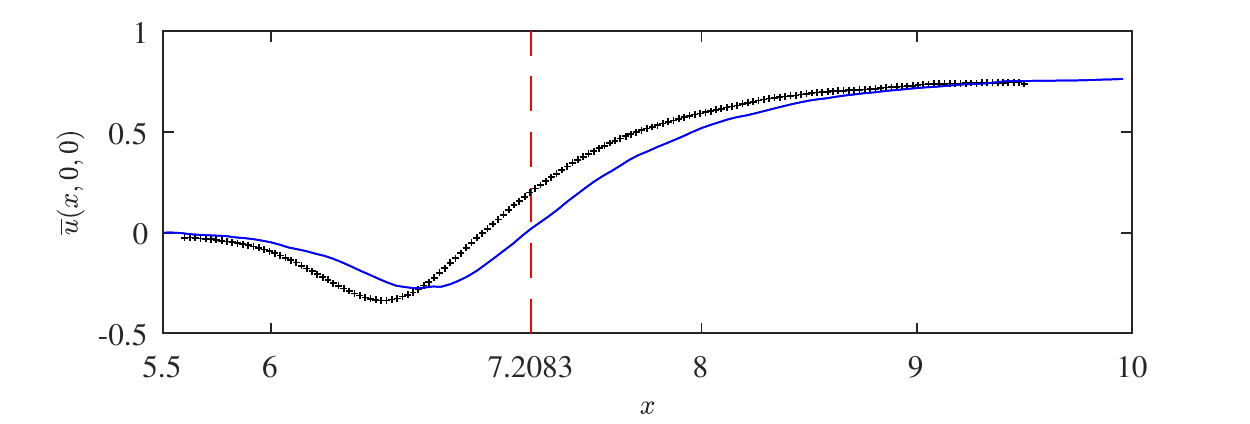}\\
~~~~~~~~~~~~~~~~~~~~~~~\\
\caption{
 Mean stream velocity component in a circular cylinder wake flow at Reynolds number $Re=3900$: (black crosses), PIV measurements \citep{parnaudeau2008experimental}; (blue line), LES used in this paper.}
\label{fig_u_bar}
\end{figure}

In the following (with the exception of \S \ref {ssec:turbulent velocity components}), non-dimensional quantities are considered, and calculated using the cylinder diameter $D$ and the inflow velocity $U_0$. Dimensionless quantities will be identified by lower-case symbols, e.g. $(x,y,z)$ for the coordinate system and $t$ for the time. In this frame reference, the inflow velocity vector at $x=0$ is $(u,v,w)=(1,0,0)$ and the cylinder is located at $(x,y,z)=(5,0,z)$.
Details of the three cases are provided in table \ref{tab:cases}. Figure \ref{fig_vorticity100} shows the spanwise vorticity component in the plane $z=0$ for the DNS at Reynolds number $Re=100$ and the  LES at Reynolds number $Re=3900$ respectively. At Reynolds number $Re=100$ and due to the quasi two-dimensional configuration of the simulation, there are only few small-scale features. Most of the energy is gathered in the large-scale vortical structures. In this regime two topos modes are sufficient to reliably describe the flow. At Reynolds number $Re=3900$, a sustained turbulence can be observed in the far wake of the cylinder and in the recirculation zone just behind the cylinder. The boundary layer on the body is laminar and transition to turbulence takes place in the shear layers. The near wake flow is mostly driven by those two shear layers \citep{Ma_2000}. Their oscillations trigger the Von Karman vortex shedding and determine the size of the recirculation area. For this wake flow regime a higher number of modes must be retained. 
Figure \ref{fig_u_bar} provides a test of the LES accuracy. There, we plotted the mean stream component of the velocity just after the cylinder in order to measure the length of the recirculation zone. To define it we here rely on the so-called bubble length. It is the distance between the base of the cylinder and the point with null longitudinal mean velocity ($\overline{u}=0$) on the centreline of the flow (y=0). Figure \ref{fig_u_bar} shows a bubble length of about $1.7$. This value is larger than experimental measurements \citep{parnaudeau2008experimental} as expected with the classical Smagorinsky subgrid model \citep{chandramouli_etal_2016}.

\section{Diffusion modes results}\label{Numerical results}
We now apply the novel POD modelling, based on a stochastic small-scale representation as presented in \S \ref{Stochastic reduced-order model with differentiable velocity}, to the cylinder flow configuration which is described in \S \ref{sec:flow configuration}. The reduced-order dynamics of the cylinder wake flow is known to be sensitive to the unresolved small-scale velocity component. 
In \S \ref{ssec:turbulent velocity components}, the small-scale energy and anisotropy are linked to small-scale diffusion modes. Assessment of the stochastic modelling is then performed in the following at the Reynolds number $Re=100$ and $Re=3900$, respectively. In \S \ref{ssec:small-scale energy}, contributions of small-scale diffusion modes to large-scale flows are described and interpreted to determine which physical mechanisms of the wake flow are concerned. In \S \ref{ssec:chronos}, we assess the performance of the subgrid term that was introduced by the stochastic representation of the small-scales by comparing the {\em chronos} trajectories to the reference.

\subsection{Estimation and decomposition of the turbulent velocity components}\label{ssec:turbulent velocity components}
It can be noticed in the decomposition (\ref{eq a span chronos}), that  the diffusion modes $\mathsfbi z_i(x)$ are $d \times d$ symmetric matrices (with $\mathsfbi z_i =0$ for $i>0$ in the stationary case) at all spatial points. They can be  diagonalized in a local orthonormal basis. Let us note however this decomposition does not ensure that  $\mathsfbi a(x,t)=\sum_{k=0}^n b_k(t) \mathsfbi z_k(x)$ is positive definite since $b_i(t)$ takes positive and negative values. In practice, though the stationary coefficient dominates largely the  other coefficients (which gives a positive definite estimation), we would have to project the variance tensor on the manifold of positive definite matrices. In the following section, to analyse the small-scale energy and anisotropy we visualize the absolute values of the eigenvalues associated with the matrix modes, $ \mathsfbi z_i$. Due  to {\em chronos } normalization, the variance tensor diffusion modes must be also normalized by  the {\em chronos}' square root eigenvalues $\sqrt{\lambda_i}$, as:
\begin{equation}
 \tilde {\mbs{a}}(x,t)
 =
 \sum_{k=0}^n \frac{b_k (t)}{\sqrt{\lambda_k}} \left ( \sqrt{\lambda_k} z_k(x) \right ) 
 \text{ and } 
\frac 1 T \int_0^T \left ( \frac{b_k}{\sqrt{\lambda_k}} \right )^2 =1 
\text{ (no sum on }k)
 .
\end{equation}
We note this normalization puts an even stronger  emphasis on the stationary dissipation zero-mode. 
Then, by \eqref{ef-adv} the corrective drift reads
\bea
 \w^\star - \w 
= \sum_{k=0}^n \frac{b_k (t)}{\sqrt{\lambda_k}}  
(\mbs v_c)_{k}
\text{ with }
(\mbs v_c)_{k}
 = 
- \frac 12 \nab \bcdot \left( \sqrt{\lambda_k} \mathsfbi z _k \right)\transp 
\text{ (no sum on }k)
 .
\eea

Before dealing with \textit{chronos} reconstruction, we propose in \S\ref{ssec:small-scale energy} a new type of POD data analysis involving the information contained in the residual velocity. Algorithm \ref{algo sum up} summarizes the steps of our data analysis, including the POD,  the diffusion modes and the corrective drift computation.

  \begin{algorithm}
    \begin{algorithmic}
      \Function{stochastic\_POD\_analysis}
      {$n,  \mbs v (\xx,t_0), ... , \mbs v (\xx,t_N) )$}

\begin{enumerate}
\item Usual POD

Resolved velocity component
\begin{flalign}
\w(\xx,t) 
&=  b^i(t) \mbs \phi_i (\xx)
.
&&
\end{flalign}


\item Optimal time step
\begin{flalign}&
\frac{1}{\Delta t} = 2 \max_{i \leqslant n}  \left ( f_{max} \left ( b_i \right ) \right ).
&& \end{flalign}


\item Diffusion modes analysis: study of the residual velocity component influence
\begin{itemize}

\item Residual velocity component
\begin{flalign}&
\mbs v - \mbs w.
&& \end{flalign}

Decomposition of the residual velocity influence

\For{$j = 0$ to ${n}$}

Component of the residual velocity influence associated with the time variability of the \textit{chronos} $b_j$ (note that $b_0 =\lambda_0=1$)
\begin{itemize}

\item Diffusion mode computation

Projection of the squared residues on the resolved \textit{chronos} $b_j$
\begin{flalign}&
\mathsfbi z _j(\xx) 
=
  \frac{  \Delta t }{N+1} \sum_{k =0}^{N} 
  \frac{b_j (t_k)}{\lambda_j} (\mbs v -\w)(\xx,t_k) ((\mbs v-\w)(\xx,t_k) )\transp 
 \text{\hspace{0.5cm}(no sum on }j)
  .
&& \end{flalign}

\item Analysis of the diffusion of the resolved velocity $\w$ by the residual velocity
\begin{itemize}
\item Local diagonalization of the symmetric matrix $\mathsfbi z_i (\xx)$
\begin{flalign}&
\sqrt{\lambda_j} \mathsfbi z _j(\xx) 
=
\mbs  P_j (\xx) \mathsfbi \Lambda^{(j)}(\xx)  \mathsfbi P _j \transp(\xx) 
 \text{\hspace{0.5cm}(no sum on }j)
 ,
&& \end{flalign}
with summation neither on $j$ nor on $p$
\begin{flalign}&
 \mathsfbi P _j (\xx) \mathsfbi P _j (\xx) \transp=\mathsfbi P _j (\xx)  \transp  \mathsfbi P _j(\xx)  = \id
\text{ and } \mathsfbi \Lambda_{pq}^{(j)}(\xx)  = \delta_{pq} \mathsfbi \Lambda_{pp}^{(j)}(\xx) 
.
&& \end{flalign}

\item Inhomogeneity of the turbulent diffusion of the resolved velocity 

(proportional to the small-scale kinetic energy)
\begin{flalign}&
\sum_{p=1}^d \left | \mathsfbi \Lambda_{pp}^{(j)}(\xx) \right |
&& \end{flalign}

\item Anisotropy of the turbulent diffusion of the resolved velocity 

(equal to the anisotropy of the small-scale kinetic energy)
\begin{flalign}&
\frac{ \max_p \left | \mathsfbi \Lambda_{pp}^{(j)} (\xx)\right | }{ \min_p \left | \mathsfbi \Lambda_{pp}^{(j)} (\xx)\right | }
 \text{\hspace{0.5cm}(no sum on }p)
 .
&& \end{flalign}

\end{itemize}

\item Corrective drift
\begin{flalign}&
(\mbs v_c)_{j} (\xx)
=
- \frac 12 \nab \bcdot \left( \sqrt{\lambda_j} \mathsfbi z _j \right)\transp (\xx) 
 \text{\hspace{0.5cm}(no sum on }j)
 .
&& \end{flalign}

\begin{itemize}
\item Vorticity of the corrective drift
\begin{flalign}&
\nab \times 
(\mbs v_c)_{j}
&& \end{flalign}

\item Divergence of the corrective drift
\begin{flalign}&
\nab \bcdot 
(\mbs v_c)_{j}
&& \end{flalign}

\end{itemize}

\end{itemize}
  \EndFor
\end{itemize}

\end{enumerate}
       \EndFunction
\caption{POD and diffusion modes data analysis}
\label{algo sum up}
\end{algorithmic}
\end{algorithm}

\subsection{Small-scale  energy density, stationarity and anisotropy}\label{ssec:small-scale energy}
The turbulent kinetic energy density (TKE) was computed by the sum of the diffusion modes eigenvalues, since small-scale TKE is represented (up to a time scale) by the norm of that tensor.
	\begin{figure}
	 \subfigure[]{\includegraphics[trim = 0mm 5mm 0mm 10mm,clip,width=\linewidth]{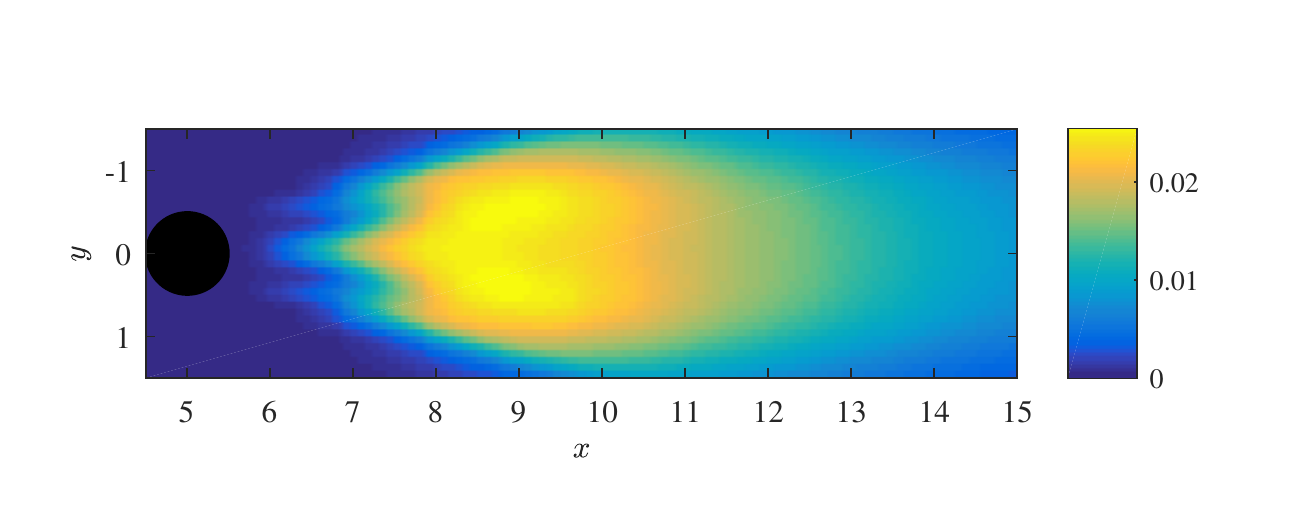} } 
	 \subfigure[]{\includegraphics[trim = 0mm 5mm 0mm 10mm,clip,width=\linewidth]{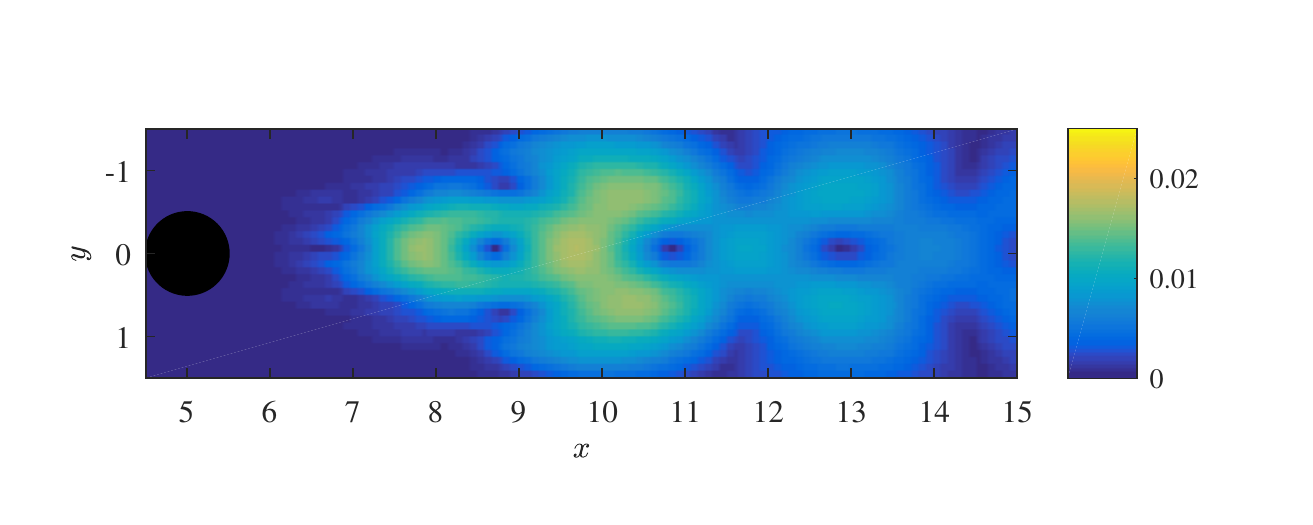} } 
	 \subfigure[]{\includegraphics[trim = 0mm 5mm 0mm 10mm,clip,width=\linewidth]{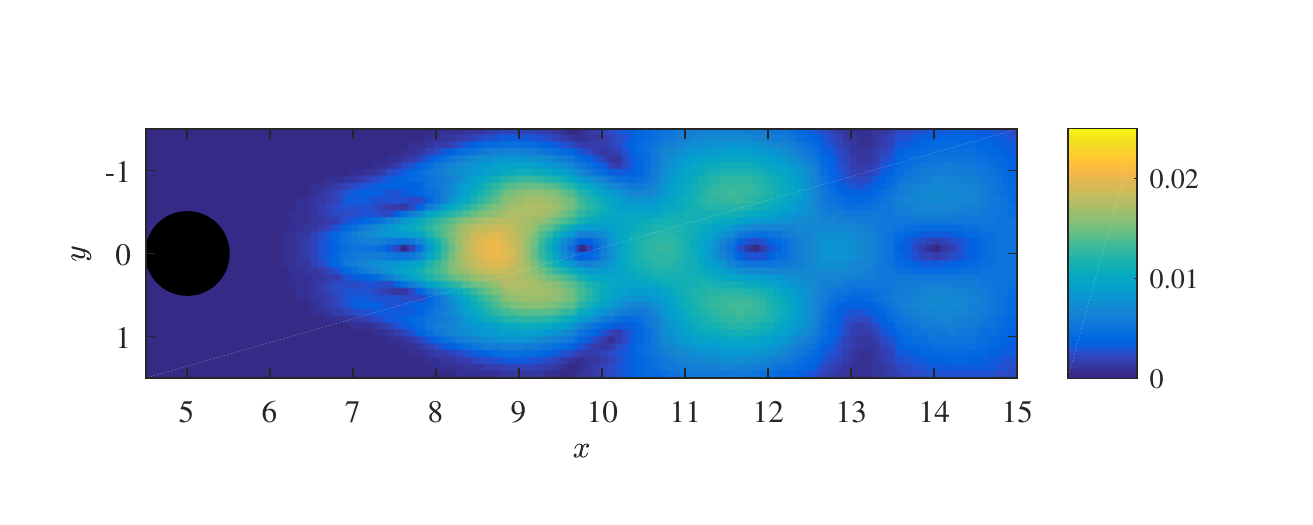} } 
	\caption{Local spectral representations of the matrix $\mathsfbi a$ in a cylinder wake flow at $Re=100$, for $n=2$ POD modes and in the plane $z=0$. Turbulent kinetic energy of the diffusion modes: (a), diffusion mode $\mathsfbi z_0$; (b), diffusion mode $\mathsfbi z_1$; (c), diffusion mode $\mathsfbi z_2$ (square root of the sum of the squared eigenvalues).}
	\label{figure z_i_2Denergy}
	\end{figure}	
The bigger the TKE, the more important  the diffusion of the resolved velocity by the small-scale velocity.
The diffusion mode energies are plotted in figure~\ref{figure z_i_2Denergy} (a-c) for the circular cylinder wake flow at $Re=100$, with two POD modes and in the plane $z=0$. 
The mode $\mathsfbi z_0$ yields regions of high TKE in the transitional region just downstream the recirculation zone , i.e. for $7.5 \leq x \leq  12$.
The other diffusion modes $\mathsfbi z_1$ and $\mathsfbi z_2$ work together since the two first \textit{chronos} $b_1$ and $b_2$ are similar up to a phase difference. These diffusion modes are twice weaker than the stationary mode $\mathsfbi z_0$. Their spatial patterns are more complex, although the large diffusion is still confined in the transitional region.

	\begin{figure}
	 \subfigure[]{\includegraphics[trim = 0mm 5mm 0mm 10mm,clip,width=\linewidth]{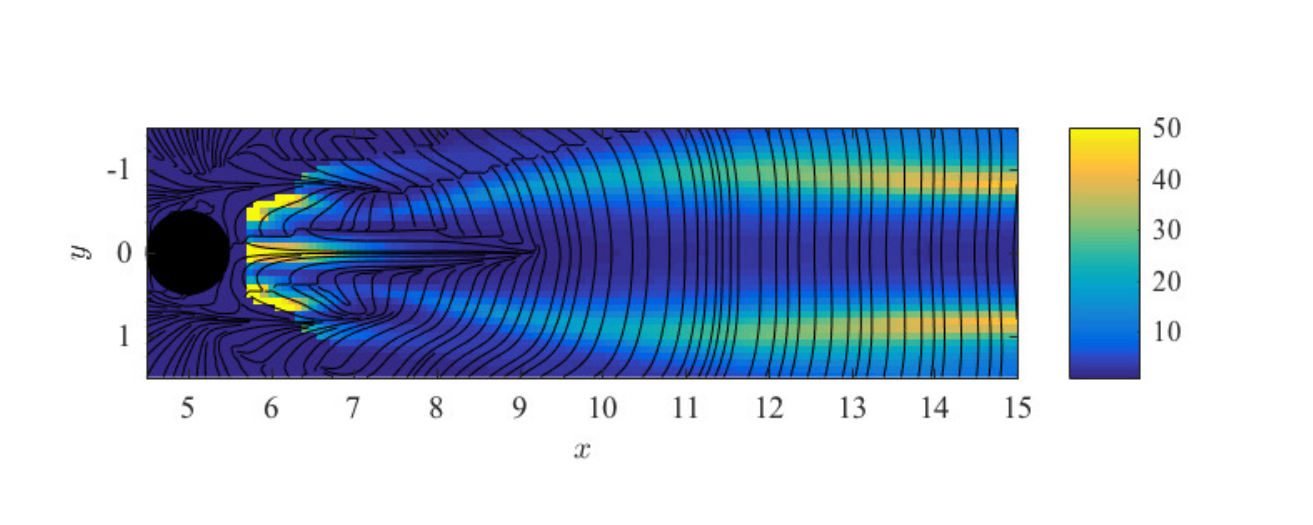}}
	 \subfigure[]{\includegraphics[trim = 0mm 5mm 0mm 10mm,clip,width=\linewidth]{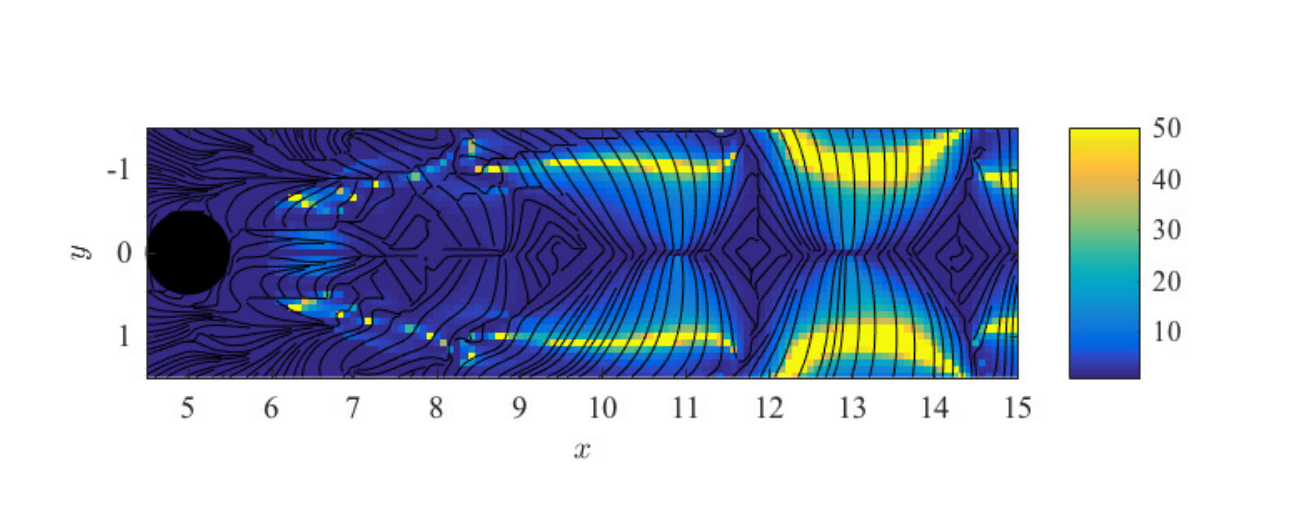}}
	 \subfigure[]{\includegraphics[trim = 0mm 5mm 0mm 10mm,clip,width=\linewidth]{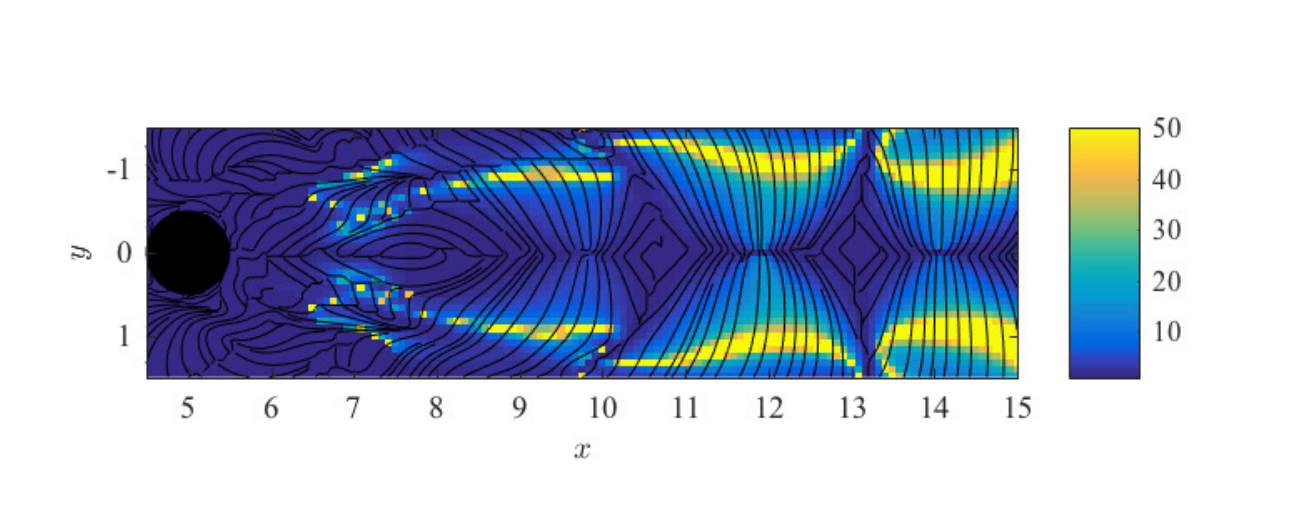}}
	\caption{Local spectral representations of the matrix $\mathsfbi a$ in a cylinder wake flow at $Re=100$, for $n=2$ POD modes and in the plane $z=0$. Small-scale anisotropy of the diffusion modes: (a), diffusion mode $\mathsfbi z_0$; (b), diffusion mode $\mathsfbi z_1$; (c), diffusion mode $\mathsfbi z_2$ (ratio of the absolute value of the largest eigenvalue to the absolute value of the smallest eigenvalue). The streamlines represent the first proper diffusion direction (i.e. the eigenvectors corresponding to the largest eigenvalues of the small-scale variance tensor).}
	\label{figure z_i_2Danisotropy}
	\end{figure}		
To measure small-scale anisotropy, we computed the ratio between largest and smallest eigenvalues, corresponding to the condition number of the local small-scale velocity variances (see Algorithm \ref{algo sum up}). The bigger this quantity the more aligned toward the first local proper direction the small-scale velocity is, i.e. the more anisotropic the small-scale velocity and the diffusion of the resolved velocity are. Figure \ref{figure z_i_2Danisotropy} shows the colormap of this quantity for the circular cylinder wake flow at $Re=100$. In regions where the unresolved velocity component  is largely anisotropic, the small-scale velocity is mainly directed  towards the eigenvector which is associated with the largest eigenvalue of the small-scale variance tensor. The small-scale component imposes a diffusion of the resolved velocity in the same direction. The streamlines in figure~\ref{figure z_i_2Danisotropy} show the principal local diffusion directions defined by the largest eigenvectors. 
The orthogonal to the streamlines would depict the directions of least diffusion of the large-scale velocity by the small-scale component. These directions can be interpreted as those of lowest small-scale uncertainty.
The streamlines clearly show the vortex formation region with the symmetric vortex rolling zone. The two pivotal locations at $y = \pm 0.5$ just before station $x = 6$ where both shear layers start to roll into vortices are precisely indicated by high values of the small-scale anisotropy. 
The centreline of the recirculation zone (just before $x=6$ up to $x=7$) is also associated with large anisotropy due to the alternation between formations of clockwise and anticlockwise vortices.
After the transitional region ($x\geq 12$), i.e. where regular and aligned vortices move downstream, the small-scale velocity anisotropy is maximum on the sides of the Karman vortex street ($x=\pm1$). This is due to the tails of the vortices visible in figure \ref{fig_vorticity100}.
This anisotropy is also visible in the non-stationary diffusion modes $\mathsfbi z_1$ and $\mathsfbi z_2$.

\begin{figure}
\subfigure[$
\frac 12 
 \nab^{\bot} \mbs \cdot \left( \mbs \nabla \mbs \cdot \mathsfbi z_0 \right )\transp$]{\includegraphics[trim = 4mm 6mm 4mm 10mm,clip,width=0.5\linewidth]{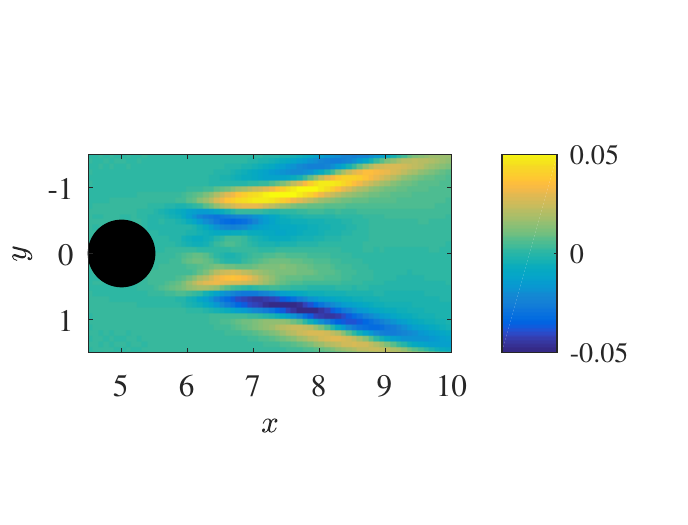}}
\subfigure[$
\frac 12 
 \nab \mbs \cdot \left( \mbs \nabla \mbs \cdot \mathsfbi z_0 \right )\transp$]{\includegraphics[trim = 4mm 6mm 4mm 10mm,clip,width=0.5\linewidth]{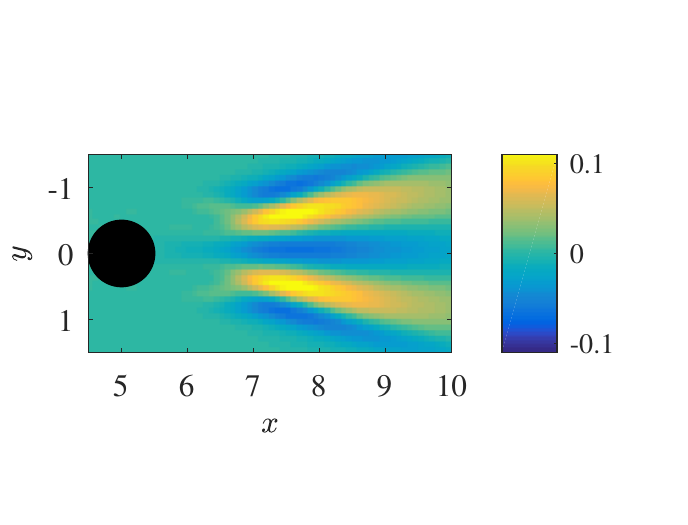}}
\subfigure[$
\frac 12 
 \nab^{\bot} \mbs \cdot \left( \mbs \nabla \mbs \cdot \mathsfbi z_1 \right )\transp$]{\includegraphics[trim = 4mm 6mm 4mm 10mm,clip,width=0.5\linewidth]{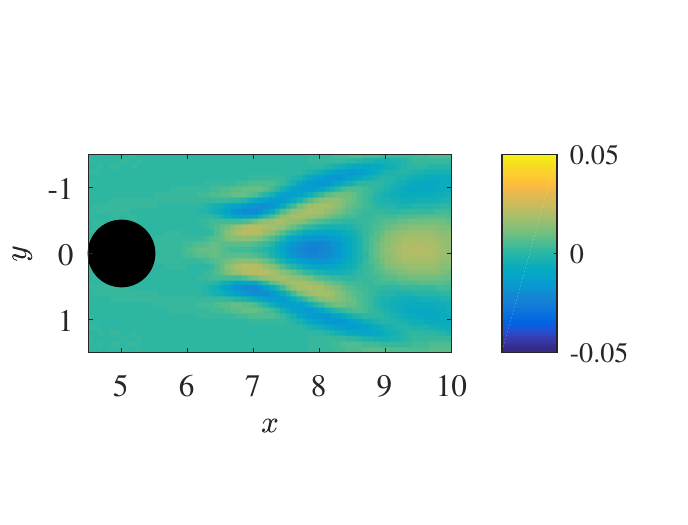}}
\subfigure[$
\frac 12 
 \nab \mbs \cdot \left( \mbs \nabla \mbs \cdot \mathsfbi z_1 \right )\transp$]{\includegraphics[trim = 4mm 6mm 4mm 10mm,clip,width=0.5\linewidth]{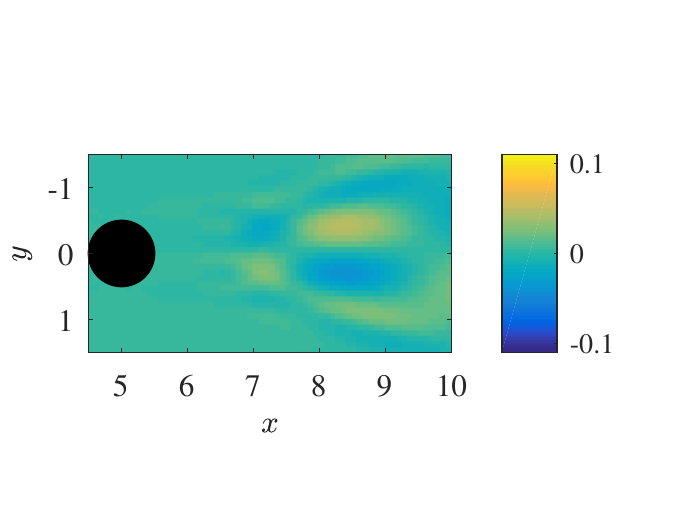}}
\subfigure[$
\frac 12 
 \nab^{\bot} \mbs \cdot \left( \mbs \nabla \mbs \cdot \mathsfbi z_2 \right )\transp$]{\includegraphics[trim = 4mm 6mm 4mm 10mm,clip,width=0.5\linewidth]{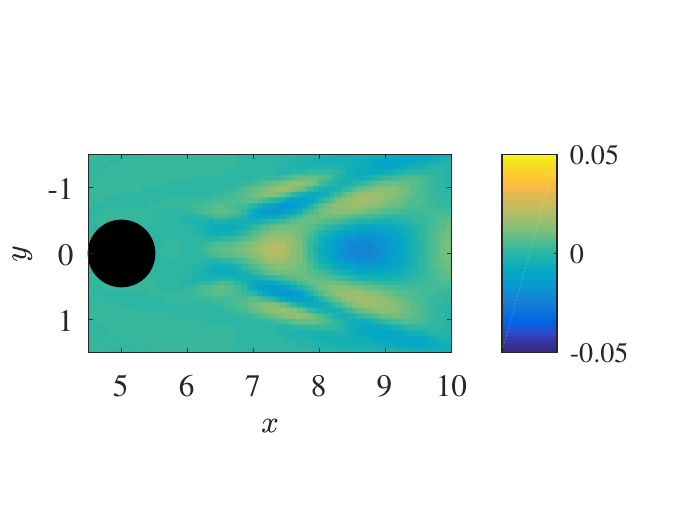}}
\subfigure[$
\frac 12 
 \nab \mbs \cdot \left( \mbs \nabla \mbs \cdot \mathsfbi z_2 \right )\transp$]{\includegraphics[trim = 4mm 6mm 4mm 10mm,clip,width=0.5\linewidth]{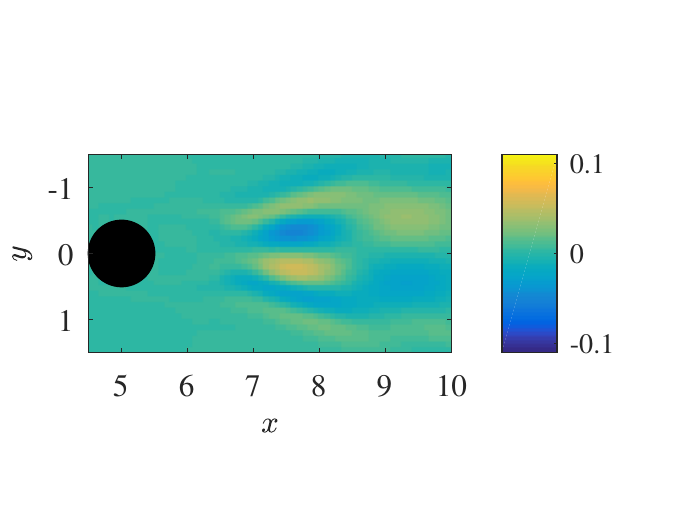}}
  \caption{
Spanwise vorticity \textit{(Left)} and divergence \textit{(Right)} of the drift correction $
	- \frac {1}{2}\left( \mbs \nabla \mbs \cdot \mathsfbi a \right )\transp$ in a cylinder wake flow at $Re=100$, for $n=2$ POD modes: \textit{(a,b)}, diffusion mode $\mathsfbi z_0$; 
\textit{(c,d)}, diffusion mode $\mathsfbi z_1$; \textit{(e,f)}, diffusion mode $\mathsfbi z_2$.
	}
\label{figure z_i_2Db}
\end{figure}	

Another interesting feature of the  small-scales stochastic representation principle concerns the emergence of the small-scale effective velocity \eqref{ef-adv}, also called drift correction, related to the variance tensor inhomogeneity (see Algorithm \ref{algo sum up} for the computation). Though at $Re=100$, this contribution is weak as the flow is well captured with only two POD modes, it is nevertheless interesting to observed the velocity component that is induced by the neglected POD modes. In figure \ref{figure z_i_2Db} (a-f) we plot the vorticity and divergence of this advection correction term \eqref{ef-adv} for the diffusion modes $\mathsfbi z_0$ to $\mathsfbi z_2$.
The small-scale vorticity induced by the neglected modes is 2 orders of magnitude weaker than the whole flow vorticity (figure \ref{fig_vorticity100}), which confirms its minor effect on the large-scale flow. However some interesting patterns emerge from these figures.
In the vorticity map associated to the diffusion mode $\mathsfbi z_0$ (figure \ref{figure z_i_2Db} a) we observe high divergence zones located at $6 \leq x \leq 7$  with $y$ close to $\pm0.5$. In these regions where both shear layers roll into vortices, the vorticity drift correction due to the unresolved modes enhances the rolling process. Just downstream, at the boundaries of the shedding zone, elongated vorticity patterns can be observed.
In the divergence map associated to the diffusion mode $\mathsfbi z_0$ (figure \ref{figure z_i_2Db} b) we observe high divergence elongated zones located at $7 \leq x \leq 9$ corresponding the trajectories of the launched vortices along which their sizes are increasing. Convergence zones are also shown at the same stations but between and on both sides of the divergence regions.
Such flow corrections, albeit weak, take place in the region of the flow where physical mechanisms that give rise to vortex shedding are active and may have significant contributions if the flow is sensitive in these regions. One interesting feature, here, is the presence of high values of vorticity, corresponding to the maximum of anisotropy, at the two pivotal locations of the shear layers rolling into vortices.
The non-stationary corrective advection is even weaker than the stationary one. The vorticity plots figure \ref{figure z_i_2Db} (c,e) unveil elongated vortices outside the recirculation zone but also relatively circular ones in the transitional region near the $x$ axis. In the divergence fields figure \ref{figure z_i_2Db} (c,e), large spots of convergence and divergence zones -- odd with respect to the $x$ axis -- appear in the end of the recirculation zone and in the transitional region.
\\

Next it is of particular interest to analyse how the proposed small-scale stochastic modelling behaves with a more turbulent wake flow. So we consider the cylinder wake flow at Reynolds number $Re=3900$.

\begin{figure}
 \centering
 \subfigure[$\sqrt{\lambda_0}tr(\mathsfbi z_0)=(0.08;0.03;0.007)$]{\includegraphics[trim = 4mm 6mm 8mm 12mm,clip,width=0.48\linewidth]{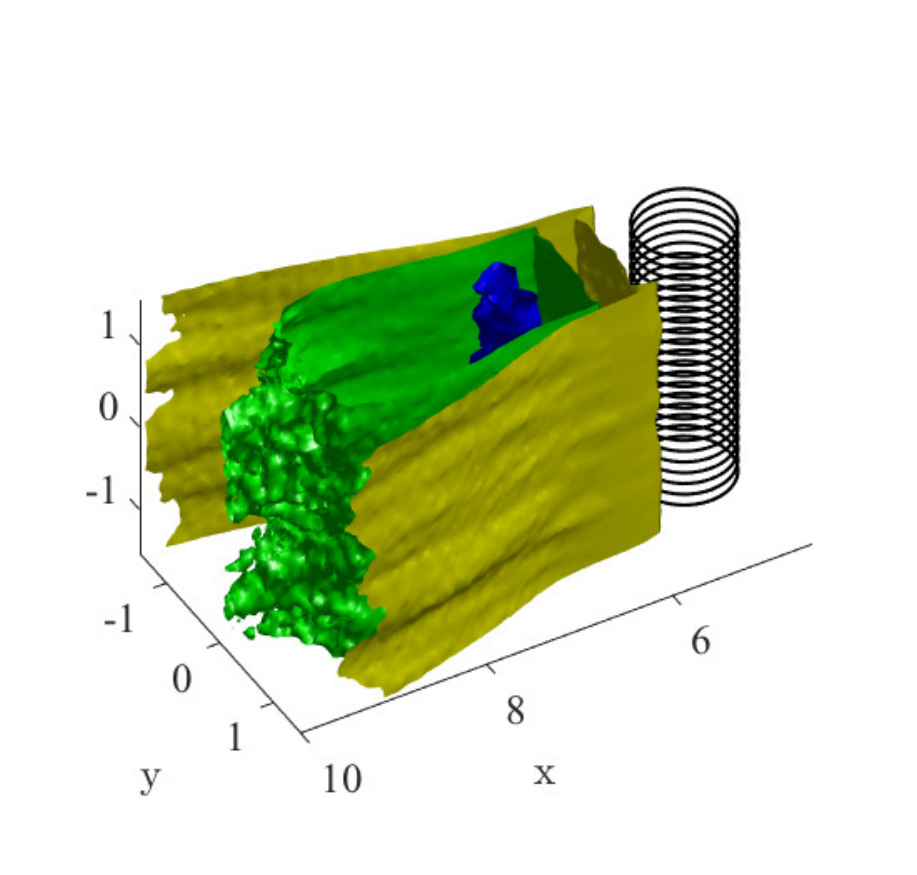}}\\
 \subfigure[$\sqrt{\lambda_1}tr(\mathsfbi z_1)=(0.03;0.007)$]{\includegraphics[trim = 4mm 6mm 8mm 12mm,clip,width=0.48\linewidth]{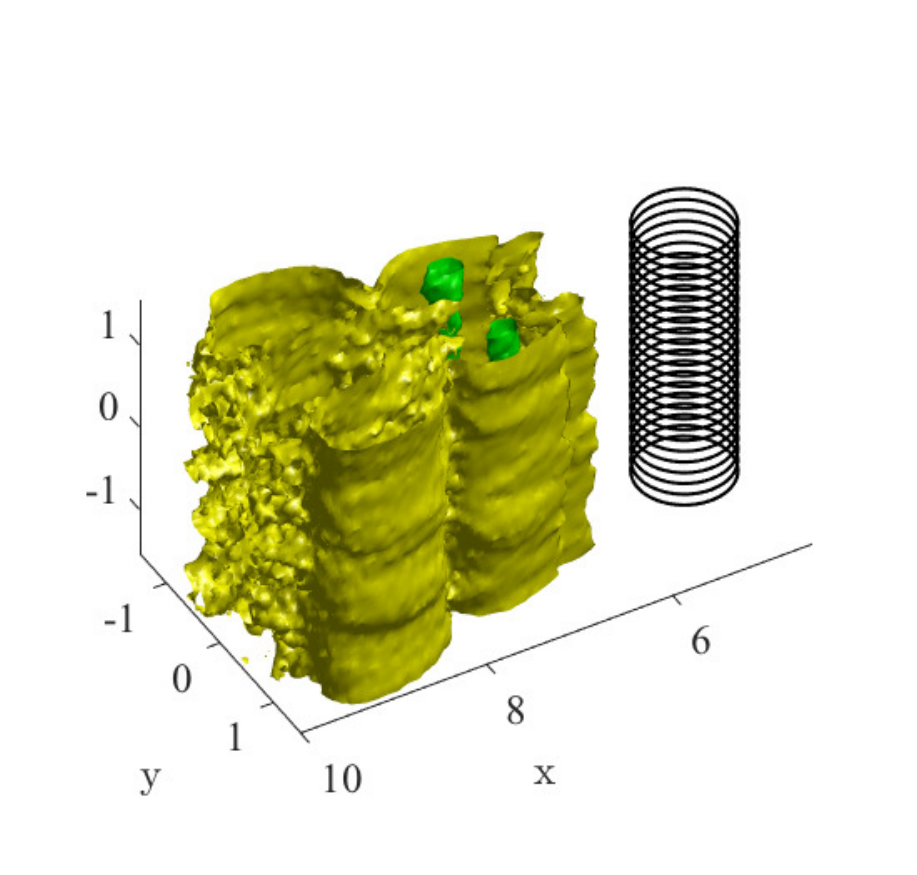}}
 \subfigure[$\sqrt{\lambda_2}tr(\mathsfbi z_2)=(0.03;0.007)$]{\includegraphics[trim = 4mm 6mm 8mm 12mm,clip,width=0.48\linewidth]{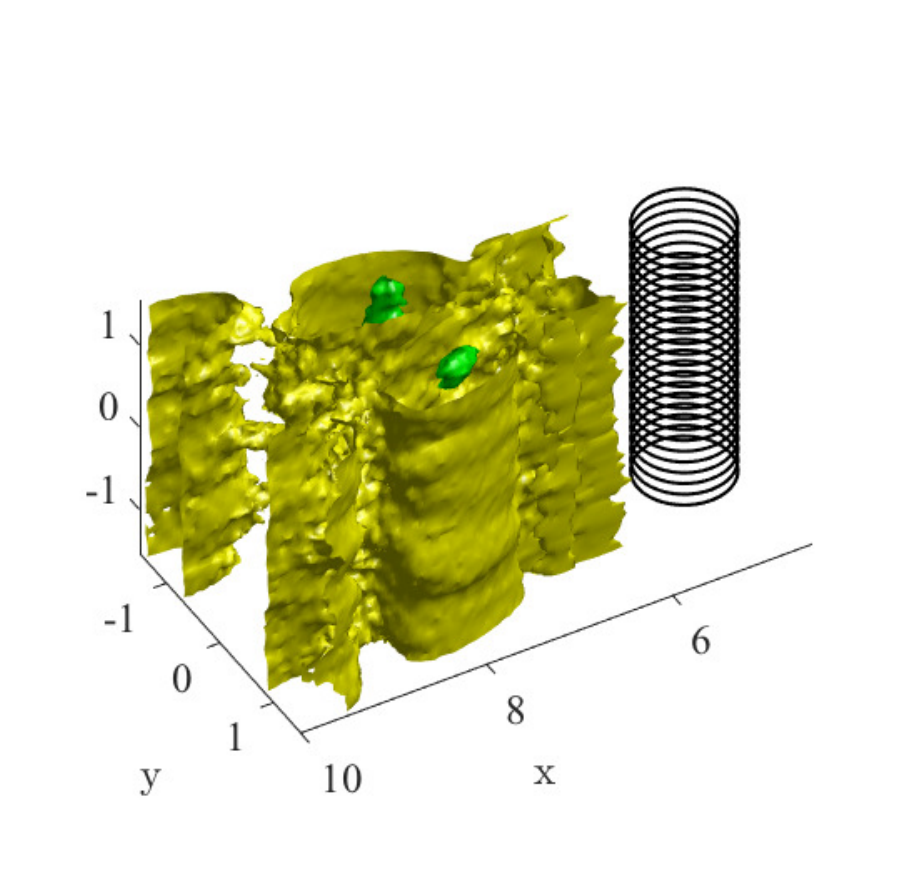}}\\
 \subfigure[$\sqrt{\lambda_3}tr(\mathsfbi z_3)=0.007$]{\includegraphics[trim = 4mm 6mm 8mm 12mm,clip,width=0.48\linewidth]{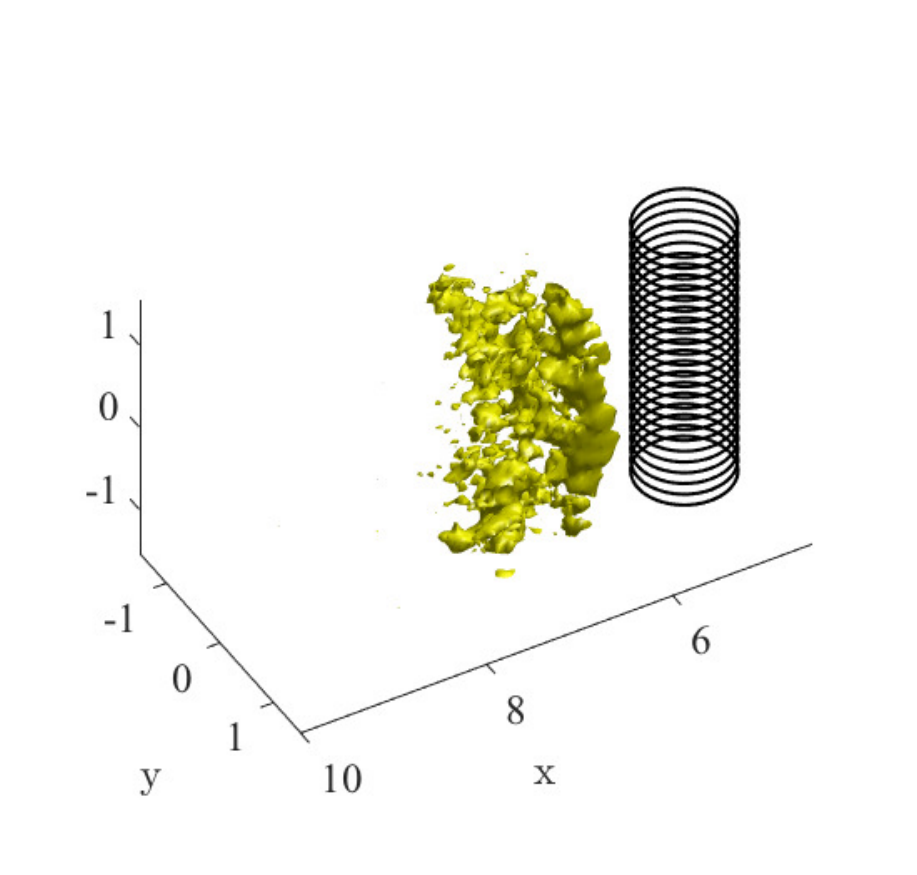}}
 \subfigure[$\sqrt{\lambda_4}tr(\mathsfbi z_4)=0.007$]{\includegraphics[trim = 4mm 6mm 8mm 12mm,clip,width=0.48\linewidth]{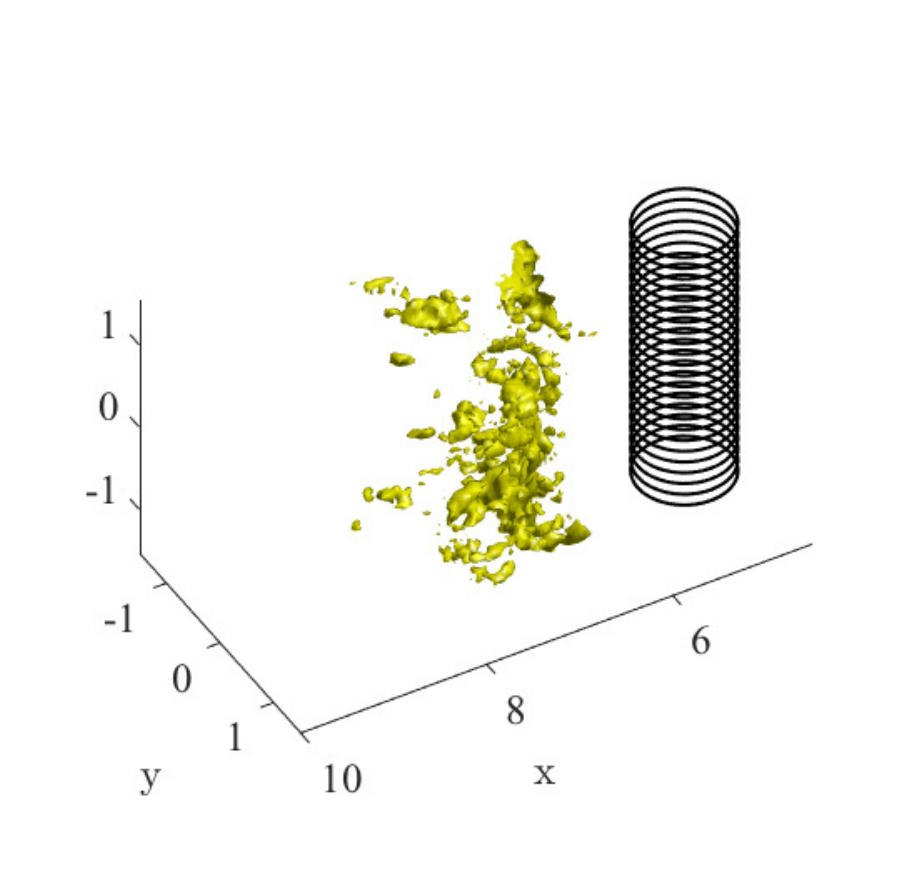}}
 \caption{Local spectral representation of the matrix $\mathsfbi a$ in a cylinder wake flow at $Re=3900$ (LES), for $n=4$ POD modes: Small-scale turbulent kinetic energy isosurfaces of the diffusion modes $\mathsfbi z_0$ to $\mathsfbi z_4$, respectively.
At places where the energy is high, the unresolved velocity and the diffusion are strong. The green isosurfaces are associated with higher values than the yellow isosurfaces ($0.007$) and the blue isosurface ($0.08$) corresponds to a higher value than the green isosurfaces ($0.03$).}
 \label{iso nrj z0}
 \end{figure}
 
Figure~\ref{iso nrj z0} is a mapping of three-dimensional iso-surfaces of the energy density for the diffusion modes $\mathsfbi z_0$, $\mathsfbi z_1$, $\mathsfbi z_2$, $\mathsfbi z_3$, and $\mathsfbi z_4$, in a cylinder wake flow at $Re=3900$. 
We observe that the turbulent energy of the diffusion zero-mode is about three times larger than for the nonstationary modes. These spatial small-scale energy distributions show that the largest magnitudes are reached at the end of the recirculation region and further downstream in the transitional region. 
 
\begin{figure}
 \centering
 \subfigure[]{\includegraphics[width=0.4\linewidth]{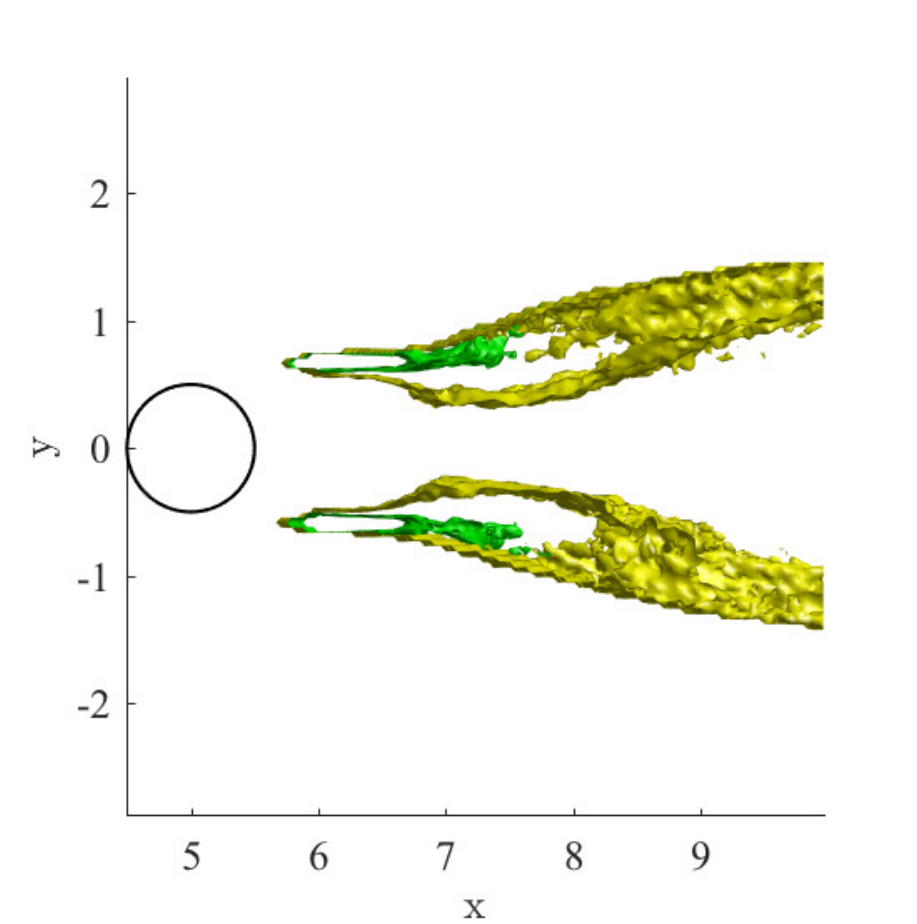}}
 \subfigure[]{\includegraphics[trim = 4mm 6mm 8mm 12mm,clip,width=0.48\linewidth]{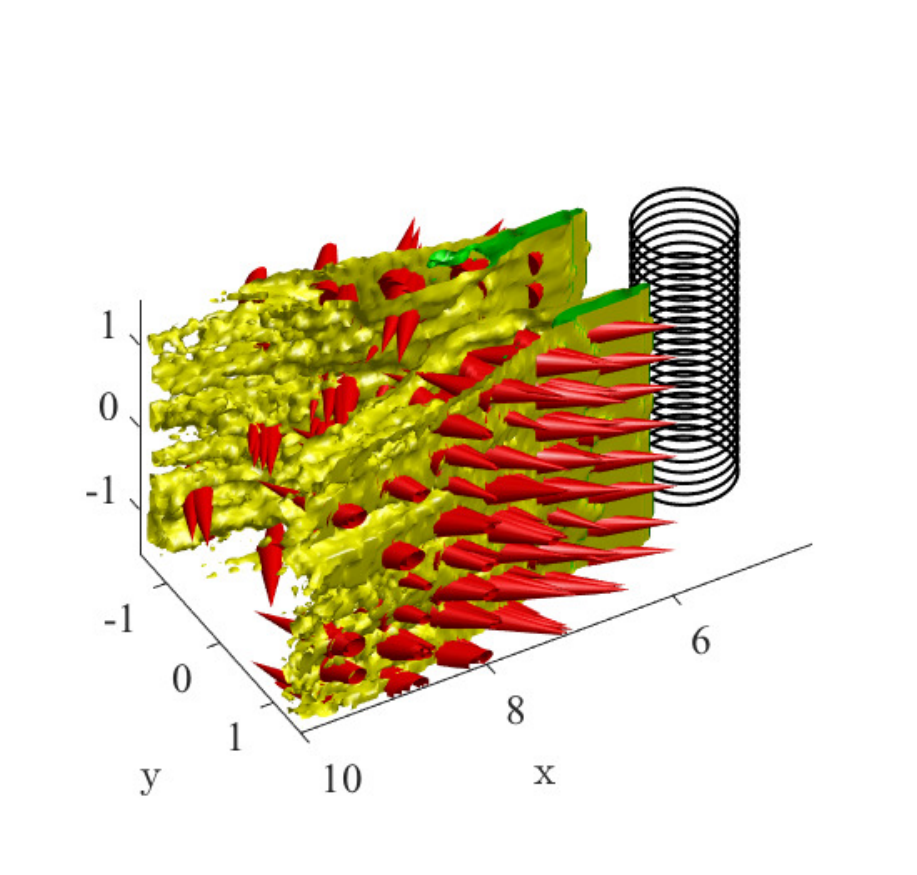}}
 \subfigure[]{\includegraphics[width=0.4\linewidth]{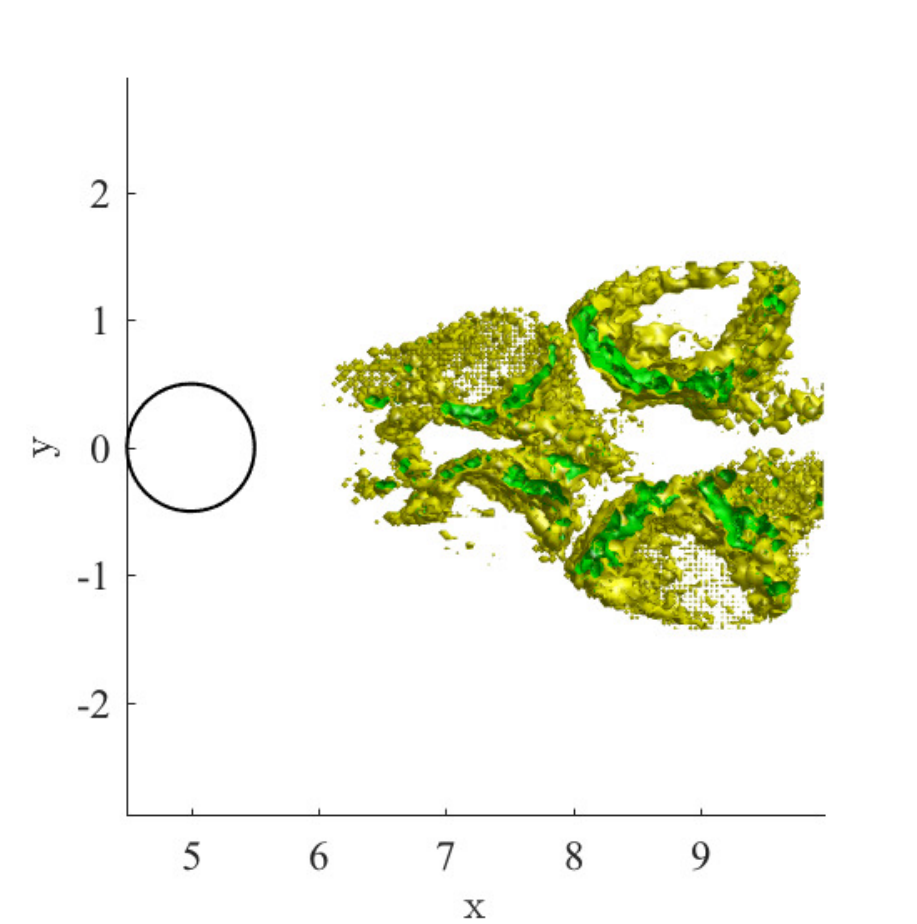}}
 \subfigure[]{\includegraphics[trim = 4mm 6mm 8mm 12mm,clip,width=0.48\linewidth]{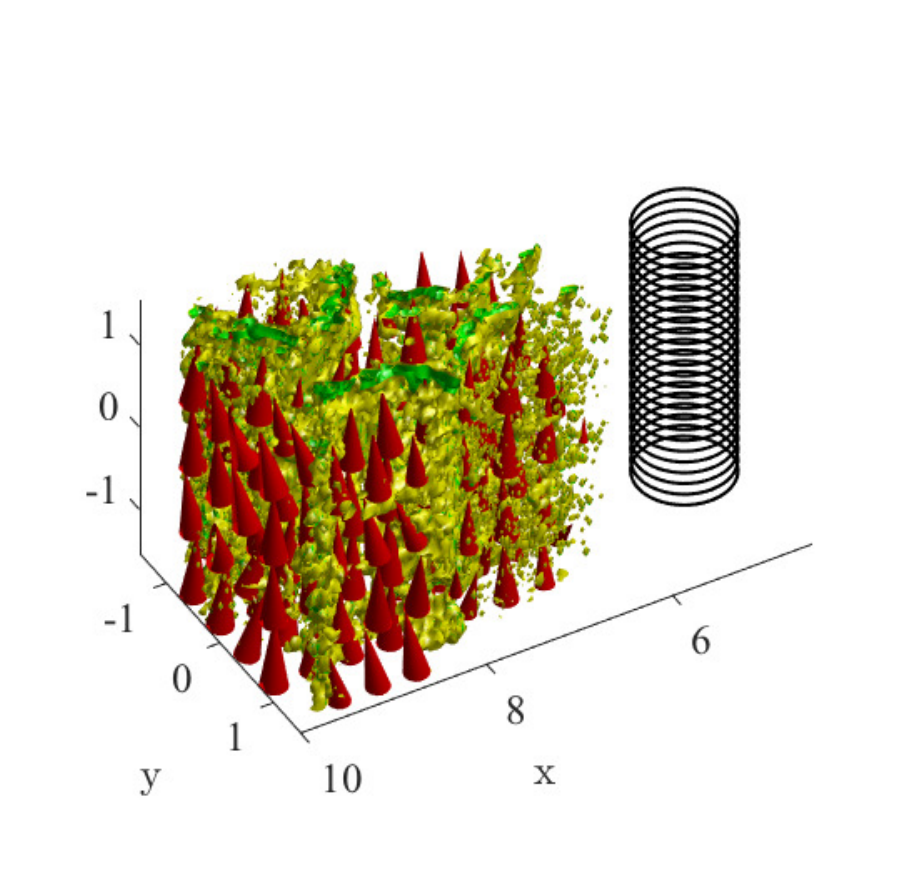}}
 \subfigure[]{\includegraphics[width=0.4\linewidth]{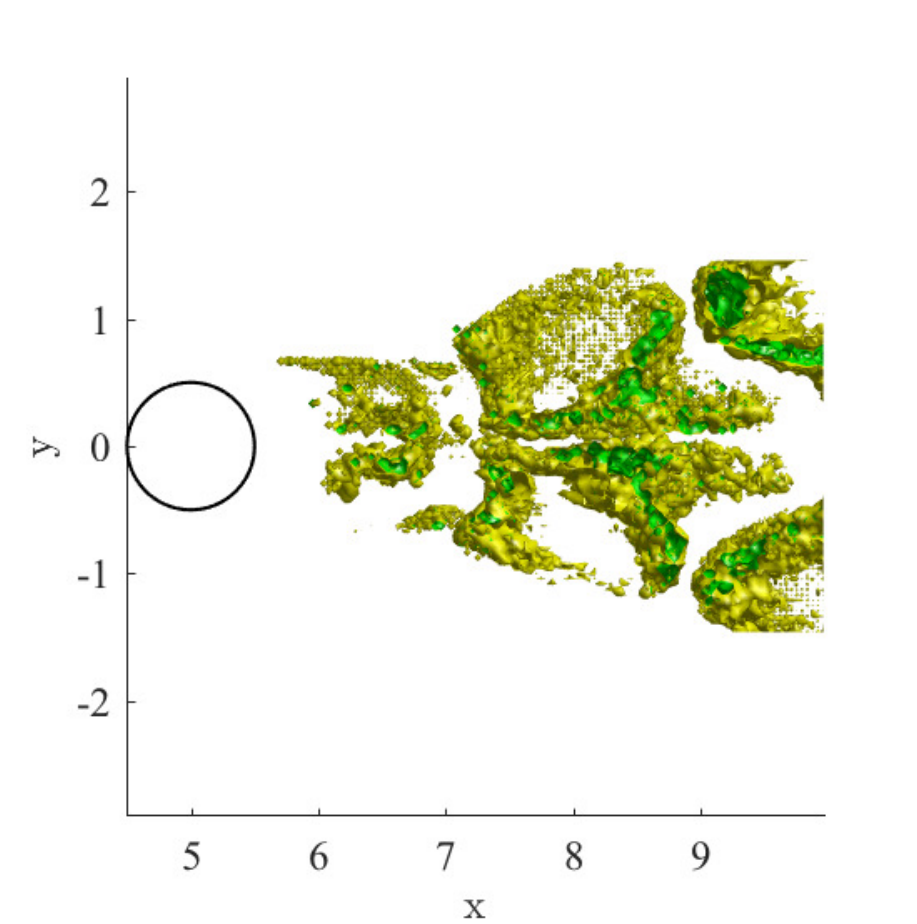}}
 \subfigure[]{\includegraphics[trim = 4mm 6mm 8mm 12mm,clip,width=0.48\linewidth]{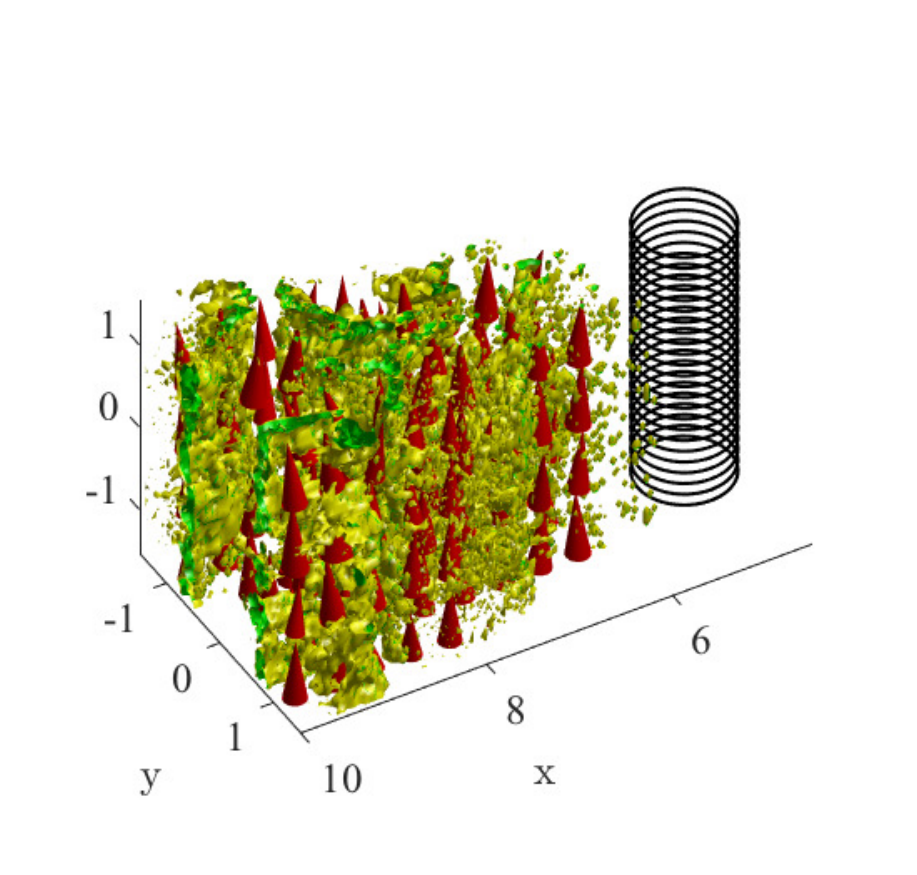}}
 \caption{Local spectral representation of the matrix $\mathsfbi a$ in a cylinder wake flow at $Re=3900$ (LES), for $n=4$ POD modes: Top (\textit{left}) and side view (\textit{right}) of the small-scale anisotropy isosurfaces of diffusion modes $\mathsfbi z_0$ (a,b) , $\mathsfbi z_1$ (c,d) and $\mathsfbi z_2$ (e,f), respectively. This is the anisotropy of both the small-scale velocity statistics and of the diffusion of the large-scale velocity.
 The green surface ($\sigma=6$) is associated with a higher anisotropy  than the yellow surface ($\sigma=3$). 
The red cones represent the preferential diffusion directions (i.e. the eigenvectors corresponding to the largest eigenvalues of the small-scale variance tensor). For the diffusion modes $\mathsfbi z_1$ \textit{(c,d)} and $\mathsfbi z_2$ \textit{(e,f)}, these direction fields are spatially smoothed for an easier visualisation.
 }
\label{iso aniso a}
\end{figure}

Let us now examine the  small-scale anisotropy spatial distribution together with the 
arrows of the principal local diffusion directions plotted in figure \ref{iso aniso a}. 
Figures \ref{iso aniso a} (a,b) indicate that the stationary component of the turbulent diffusion of the resolved velocity is isotropic (i.e. shows a low anisotropy) in the centre of the Karman vortex street and further downstream whereas this stationary diffusion is anisotropic in the sides of the recirculation zone and of the Karman vortex street. There, the turbulent diffusion mostly acts along the plane perpendicular to the cylinder.
The anisotropy maximums (highlighted by green surfaces in figures \ref{iso aniso a} a-b) reveal the two pivotal regions where the shear layers start to roll into vortices.

For $\mathsfbi z_1$ and $\mathsfbi z_2$ diffusion modes the structures of the non-stationary anisotropy are more complex (see figures \ref{iso aniso a} c,e), although  large values are mainly confined inside the Karman vortex street. 
The associated non-stationary turbulent diffusion is preferentially in the spanwise direction. This behaviour can be related to the three-dimensionalization of the wake flow especially taking place along the spanwise direction in the transitional region.
Our analysis gives an additional comprehension of this phenomena. In particular, the associated spanwise turbulent viscosity coefficient is found to be non-stationary.

\begin{figure}
 \centering
 \subfigure[$
\frac 12 
 |\nab\times \left( \mbs \nabla \mbs \cdot \mathsfbi z_0 \right )\transp|=0.2$]{\includegraphics[trim = 4mm 6mm 8mm 12mm,clip,width=0.48\linewidth]{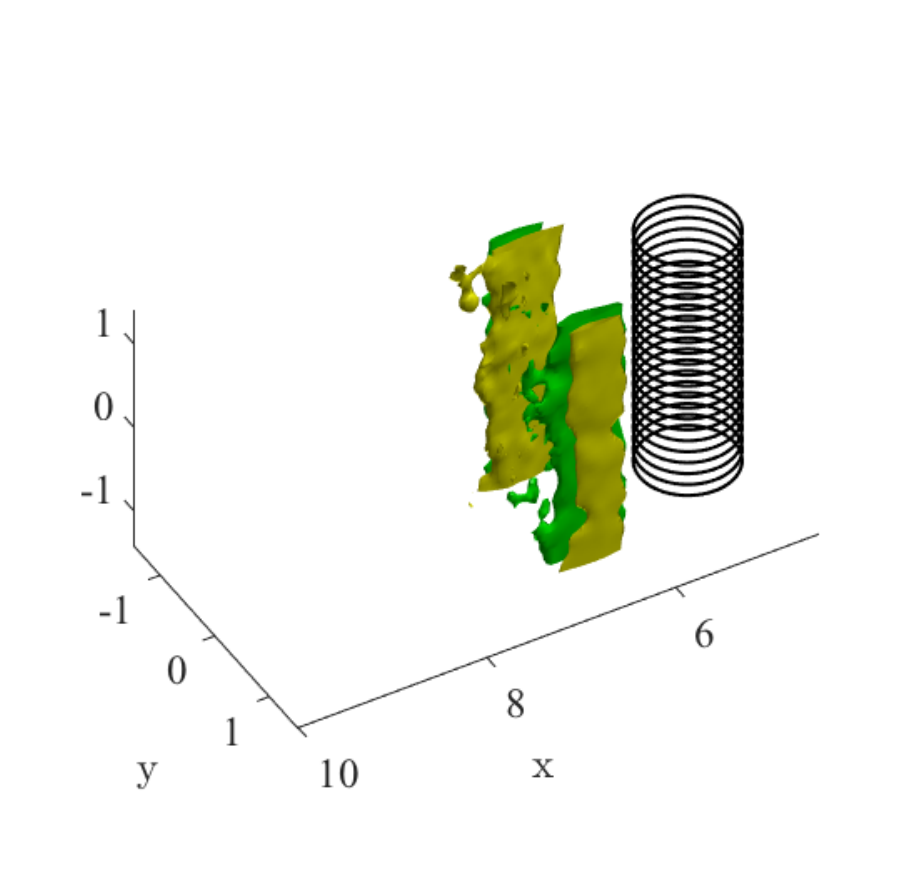}}
 \subfigure[$
\frac 12 
 |\nab  \mbs \cdot \left( \mbs \nabla \mbs \cdot \mathsfbi z_0 \right )\transp|=0.15$]{\includegraphics[trim = 4mm 6mm 8mm 12mm,clip,width=0.48\linewidth]{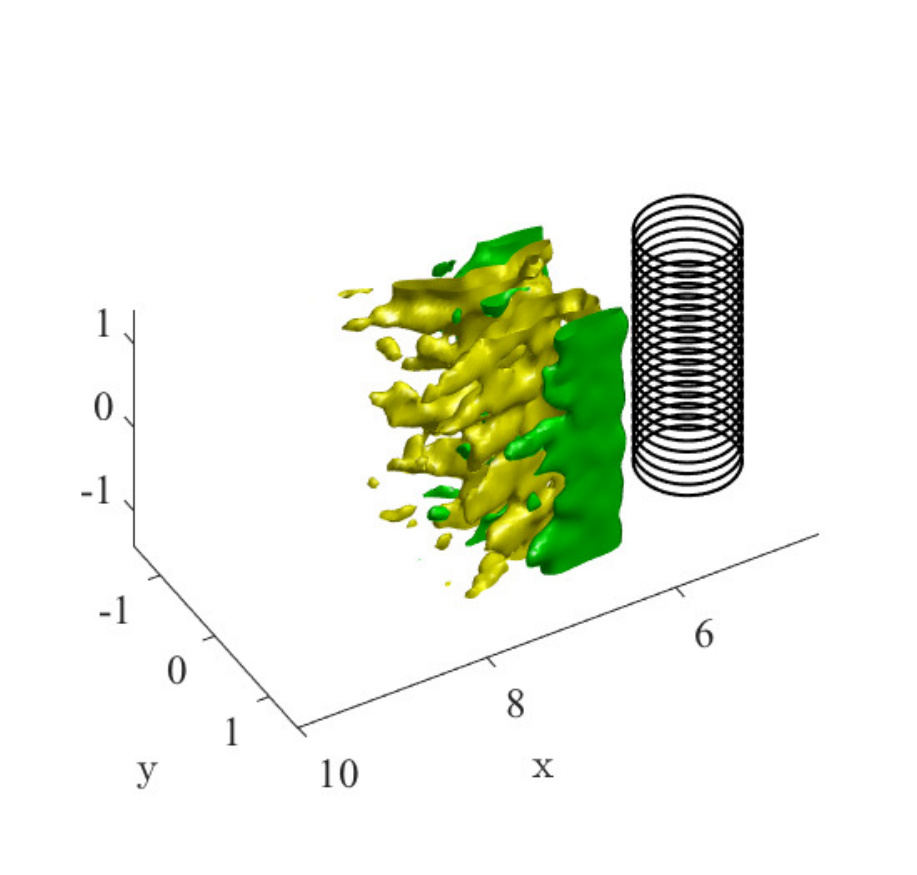}}
 \subfigure[$
\frac 12 
 |\nab\times \left( \mbs \nabla \mbs \cdot \mathsfbi z_1 \right )\transp|=0.08$]{\includegraphics[trim = 4mm 6mm 8mm 12mm,clip,width=0.48\linewidth]{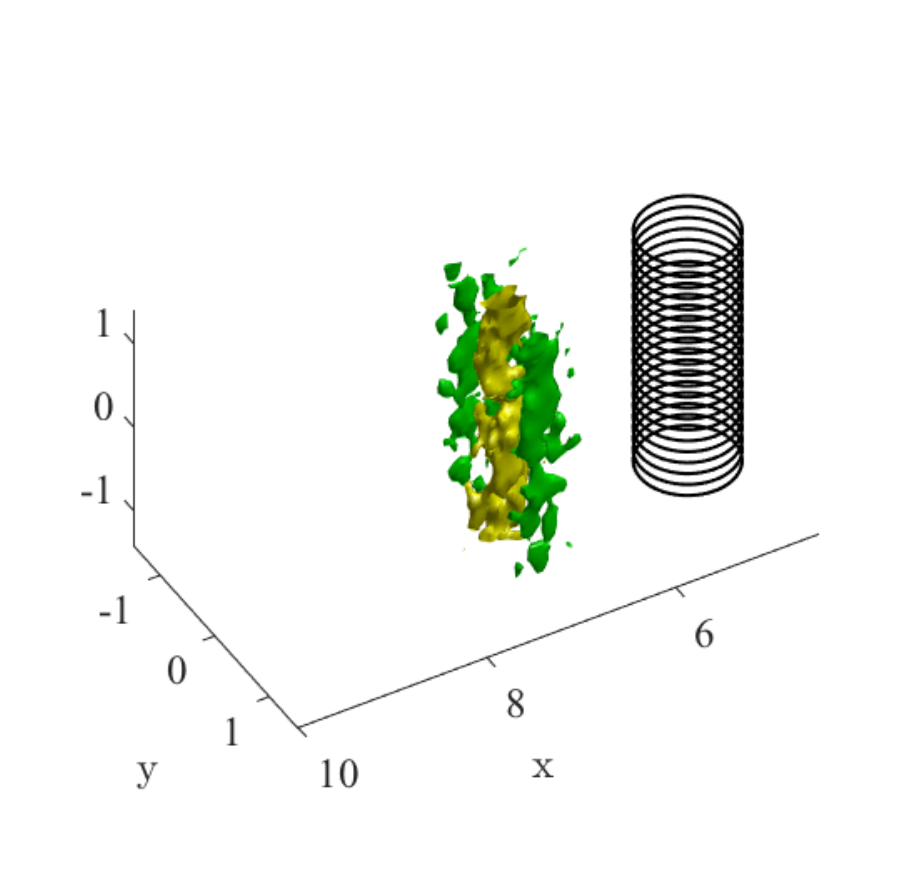}}
 \subfigure[$
\frac 12 
 |\nab \mbs \cdot \left( \mbs \nabla \mbs \cdot \mathsfbi z_1 \right )\transp|=0.05$]{\includegraphics[trim = 4mm 6mm 8mm 12mm,clip,width=0.48\linewidth]{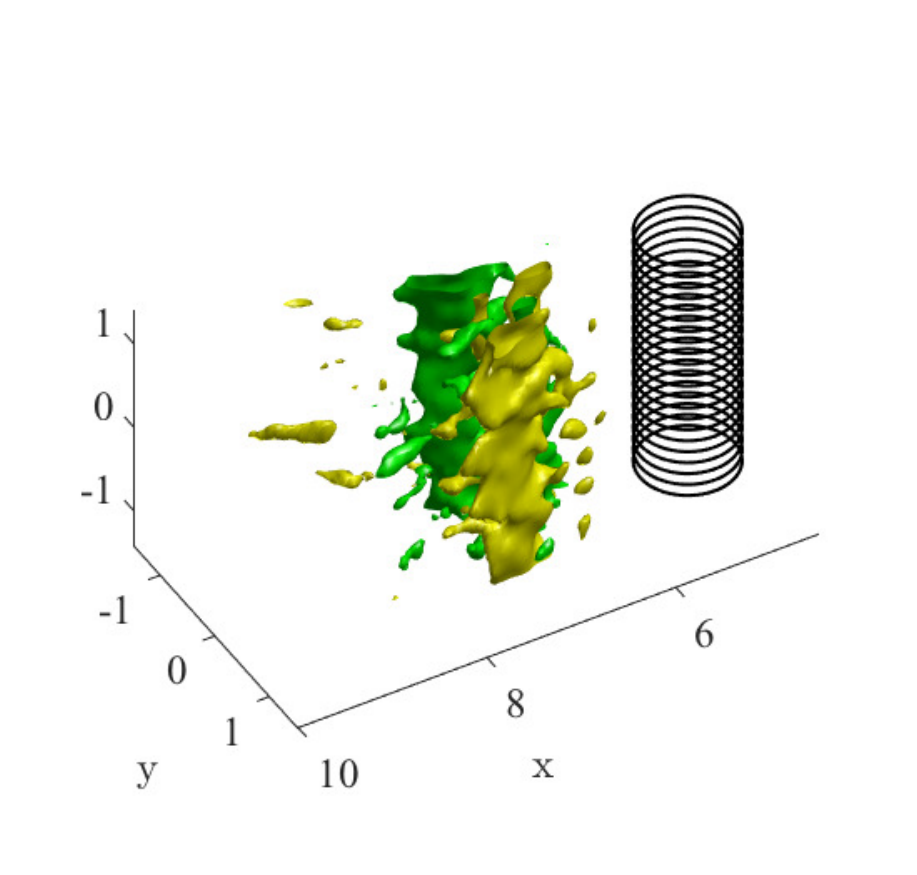}}
 \subfigure[$
\frac 12 
 |\nab\times \left( \mbs \nabla \mbs \cdot \mathsfbi z_2 \right )\transp|=0.0.08$]{\includegraphics[trim = 4mm 6mm 8mm 12mm,clip,width=0.48\linewidth]{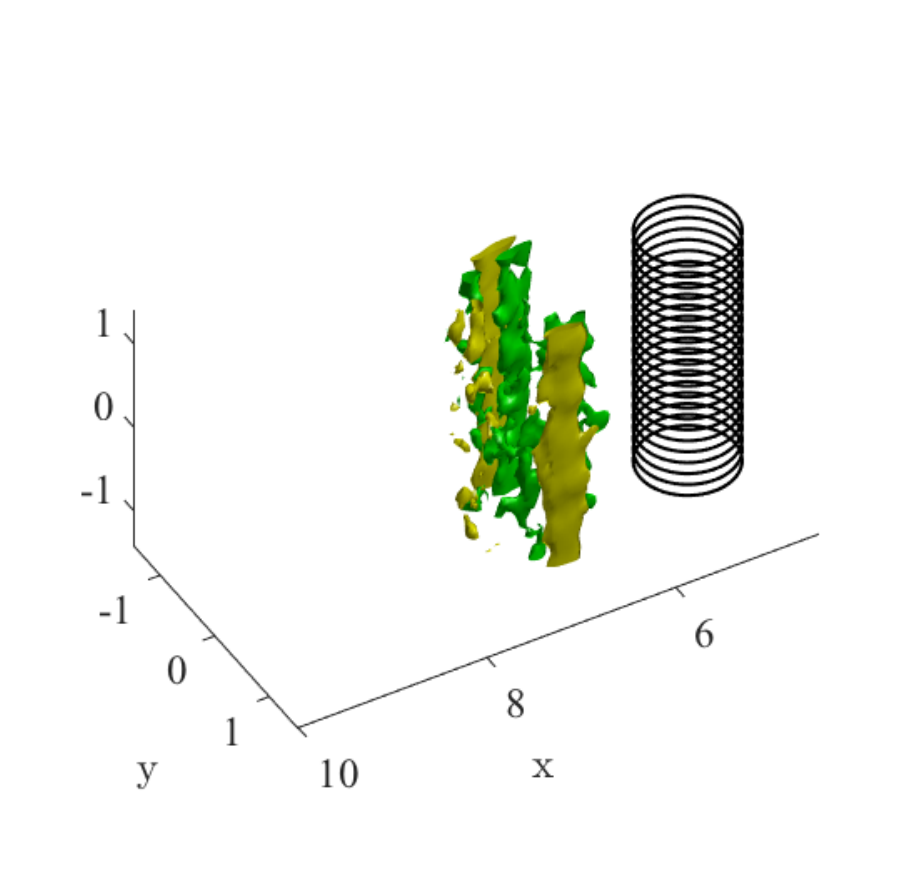}}
 \subfigure[$
\frac 12 
 |\nab \mbs \cdot \left( \mbs \nabla \mbs \cdot \mathsfbi z_2 \right )\transp|=0.05$]{\includegraphics[trim = 4mm 6mm 8mm 12mm,clip,width=0.48\linewidth]{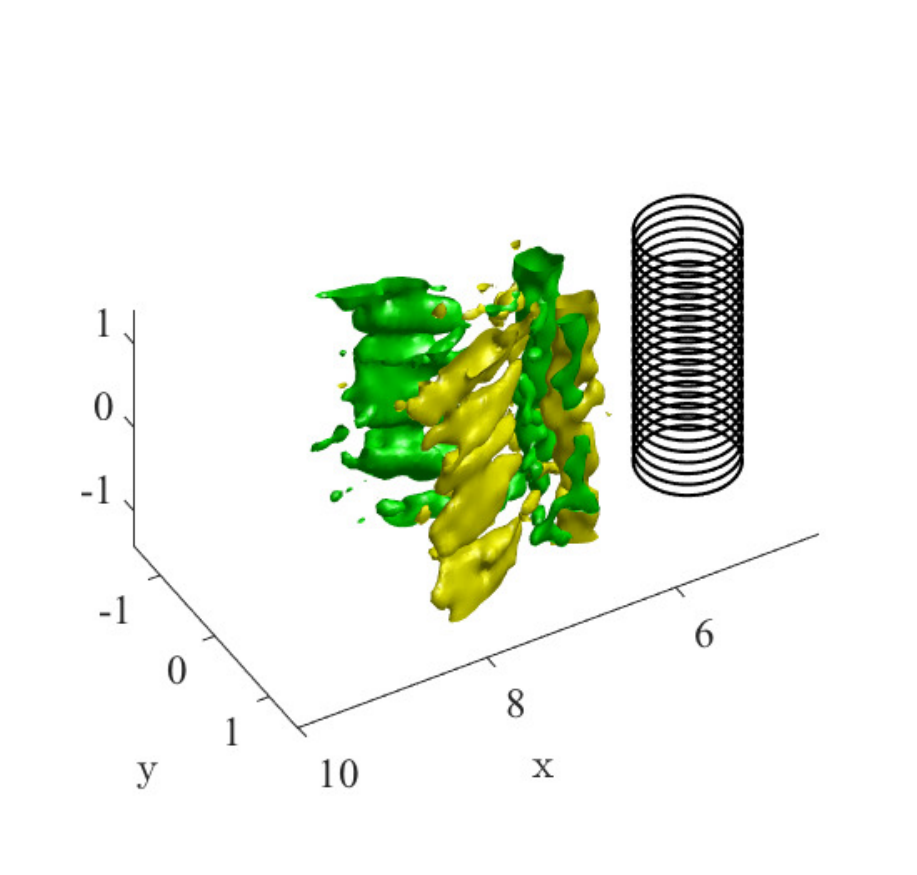}}
 \caption{Vorticity \textit{(Left)} and divergence \textit{(Right)} isosurfaces of the drift correction $
	- \frac {1}{2}\left( \mbs \nabla \mbs \cdot \mathsfbi a \right )\transp$ in a cylinder wake flow at $Re=3900$, for $n=4$ POD modes: \textit{(a,b)}, Diffusion mode $\mathsfbi z_0$; 
(c,d), Diffusion mode $\mathsfbi z_1$; (e,f), Diffusion mode $\mathsfbi z_2$.
	In the left column, the green iso-surfaces are associated with a vorticity vector aligned downward whereas the yellow iso-surfaces are associated with a vorticity vector aligned upward. In the right column the green surfaces stand for areas with iso-negative divergence (convergent zone) whereas the yellow iso-surfaces correspond to a positive divergence (divergent zone).}
\label{iso effective drift vort}
\end{figure}

 In figures \ref{iso effective drift vort} (a,b), respectively the vorticity norm and the divergence of the stationary drift correction  are plotted. Like for the $Re=100$ case, stationary vorticity and divergence corrections are observed near the shear layers, outside and at the end of the recirculation zone, and just downstream in the launching area.
Here again, high magnitudes of vorticity just downstream the maximum of anisotropy are associated with the two pivotal regions of the shear layers rolling into vortices.
Nevertheless, the corrective vorticity is here stronger (one order of magnitude weaker than the global vorticity field).
Other corrective vorticity structures can be observed at the end of the recirculation zone on the non-stationary modes $\mathsfbi z_1$ and $\mathsfbi z_2$. The vorticity involved in those modes is about twice weaker than for the stationary diffusion mode. We identify $3$ and $4$ spanwise vortices in the corrective vorticity fields $\tfrac 12 
 \nab  \times \left( \mbs \nabla \mbs \cdot \mathsfbi z_1 \right )\transp$ and $\tfrac 12 
 \nab  \times \cdot \left( \mbs \nabla \mbs \cdot \mathsfbi z_2 \right )\transp$ respectively.
As in the laminar case, divergence structures can be observed at the end of the recirculation zone and just downstream in the launching area. Again, these structures are odd with respect to the $x$ axis. Along the spanwise direction, periodic structures appear. Downstream no significant corrective velocities structures are observed, rather indicating a homogeneous character of the small-scale velocity and of the turbulent diffusion.

   The diffusion modes analysis developed in the present paper identifies critical regions of the wake flow: the anisotropy mainly exhibits the pivotal location of the shear layers which are associated with large-scale vorticity corrections by the small-scale unresolved velocity and large-scale divergence corrections also take place in the vortex formation zone. 
For the wake flow considered, 
 the results indicate that far enough downstream 
and outside the Karman vortex street boundaries
 an eddy viscosity assumption is likely to be valid. However in the near wake  or close to the Karman vortex street boundaries, such an assumption is too strong and corrective advection effects as well as structured energy dissipation effects must be taken into account. 
These findings support the recent results of \citet{chandramouli_etal_2016} who demonstrated the significant contributions of such novel stochastic small-scale modelling in the context of coarse-grid large eddy simulation of a wake flow.   

\subsection{Chronos reconstruction}\label{ssec:chronos}
In this section we aim at assessing the performance of the subgrid term introduced by the stochastic representation of the small-scales. We compare the {\em chronos} trajectories that were directly reconstructed from the reduced-order dynamical system \eqref{avant derniere eq dbi} to the observed {\em chronos}. Let us note that almost perfect long time trajectories could be recovered through data assimilation strategies \citep{Dadamo-JOT07,Artana-JCP12,Cordier_etal_2013,protas2015optimal}. However with such techniques it would be difficult to identify the intrinsic role of the subgrid scheme compared to a least-squares adaption of all the dynamics coefficients along the whole sequence.
We therefore prefer to rely on a direct reconstruction strategy in which no identification, like least-squares or data-assimilation estimation procedures of the dynamical coefficients, is introduced. Note that such a direct reconstruction requires an additional stabilization, like a closure model, to ensure the long-term boundedness of the solution. In our model, the stabilization is inherited from the rigorously derived subgrid terms based on the variance of the unresolved velocity component. In the following, we compare temporal reconstructions obtained with our stochastic Galerkin model to those provided by different deterministic POD models.

The results presented so far on the diffusion mode analysis did not necessitate any knowledge of the Reynolds number to compute both the diffusion and the drift correction of the large-scale by the unresolved small-scale.
We now turn to reduced-order dynamical systems which, in contrast, require the Reynolds value.

The modes energy mean, $\lambda_i$, and the \textit{topos}, $\bphi_i$, are computed from the whole sequence of snapshots ($N=10,000$ for $Re=100$ and $N=1,460$ for $Re=3900$). As for the initial condition, we used the referenced values of the \textit{chronos}, denoted $b_i^{ref}$, computed directly from the snapshots covariance diagonalization. Then, regarding the \textit{chronos} spectra, an optimal time sub-sampling $\Delta t$ is chosen, as explained in \S \ref{Choice of the time step}. Afterwards, using the residual velocity and the \textit{chronos}, the variance tensor, ${\mathsfbi  a}$, is estimated. 
The coefficients of the reduced-order dynamical system of \textit{chronos} \eqref{avant derniere eq dbi} are directly computed using discrete derivation schemes. The \textit{chronos} trajectories are simulated with a $4$-th order Runge-Kutta integration method, with $b^{ref}(t=0)$ as initial condition and $\Delta t/10$ as simulation time step.

	\begin{figure}
	\centering
	\includegraphics[width=\linewidth]{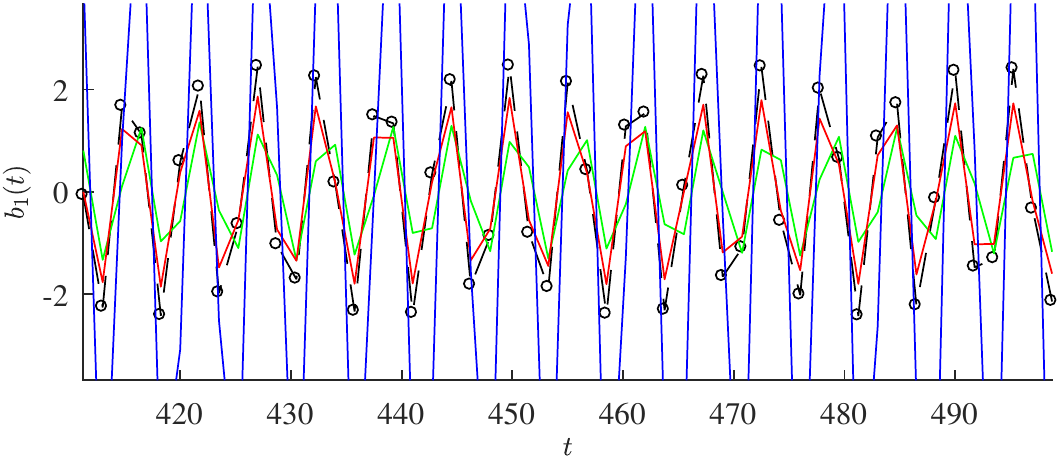}
	\includegraphics[width=\linewidth]{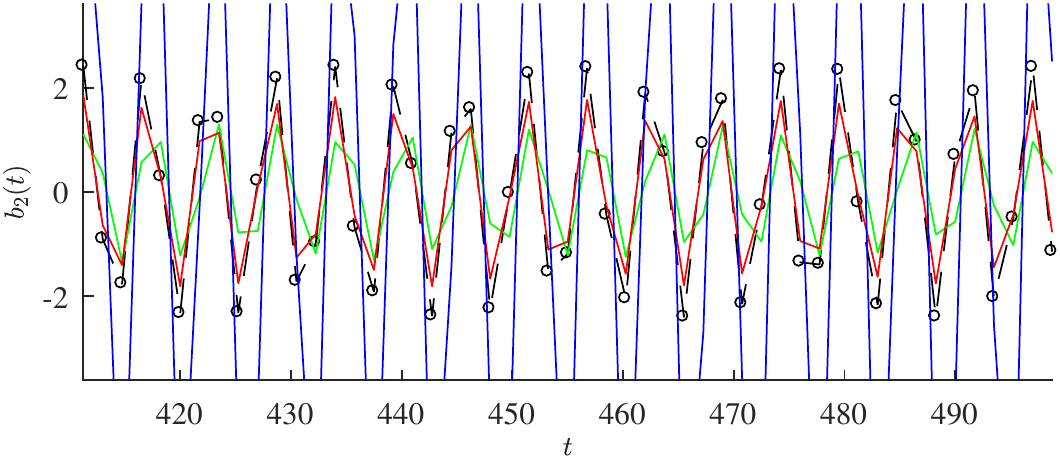}
	\caption{Reconstruction of the first two \textit{chronos} for a wake flow at Reynolds number $Re=100$, with $n=2$ POD modes and with a stationary variance tensor: (black dots), observed references; (blue line), standard POD-Galerkin; (red line), proposed stochastic representation; (green line), modal eddy viscosity reduced-order model. The initial condition, at $t=0$, is identical for all methods.
	}
	\label{plot_2D_2m}
	\end{figure}
	
	\begin{figure}
	\centering
	\includegraphics[width=0.45\linewidth]{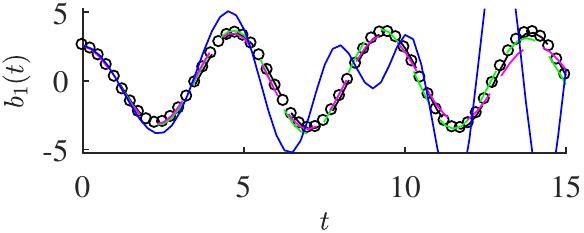}
	\includegraphics[width=0.45\linewidth]{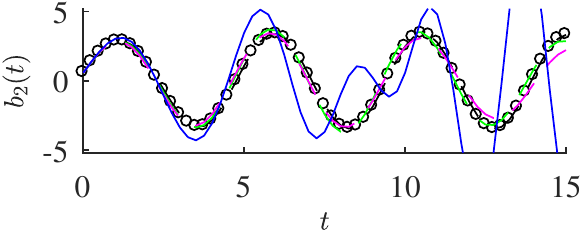}
	\includegraphics[width=0.45\linewidth]{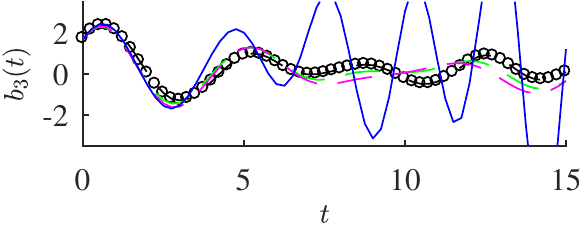}
	\includegraphics[width=0.45\linewidth]{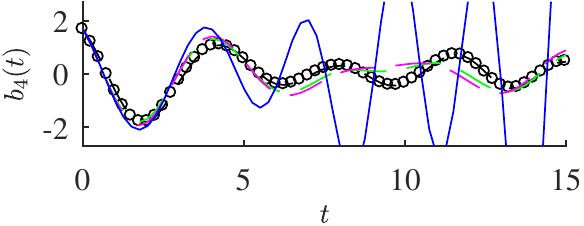}
	\includegraphics[width=0.45\linewidth]{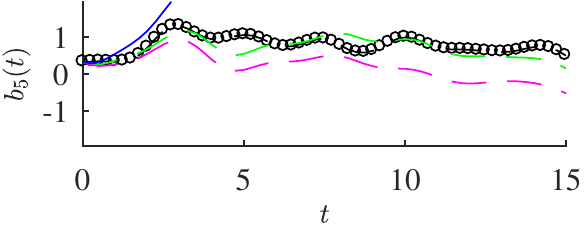}
	\includegraphics[width=0.45\linewidth]{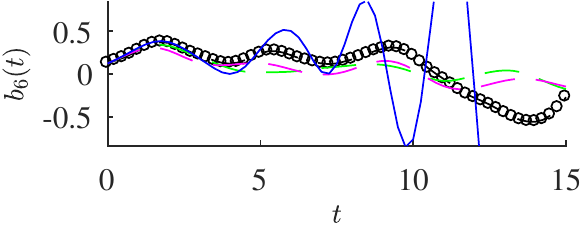}
	\includegraphics[width=0.45\linewidth]{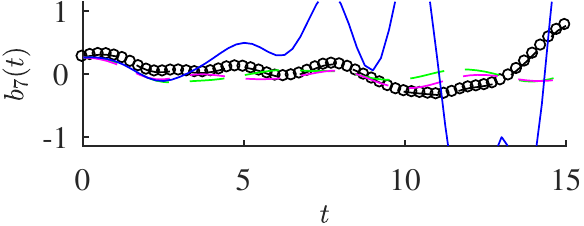}
	\includegraphics[width=0.45\linewidth]{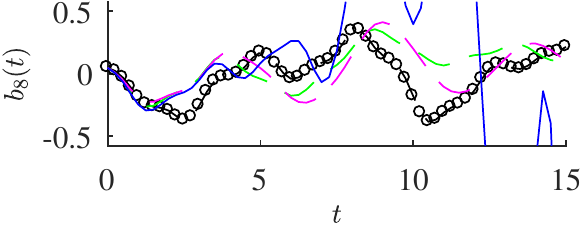}
	\includegraphics[width=0.45\linewidth]{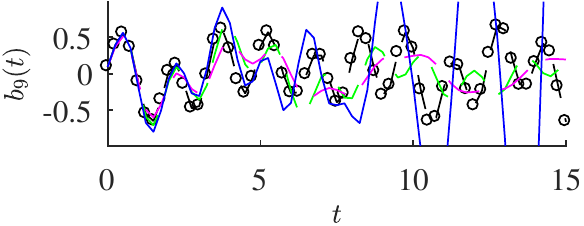}
	\includegraphics[width=0.45\linewidth]{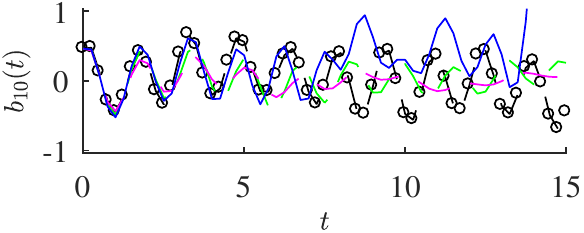}
	\includegraphics[width=0.45\linewidth]{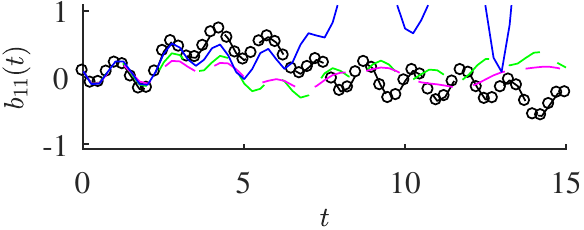}
	\includegraphics[width=0.45\linewidth]{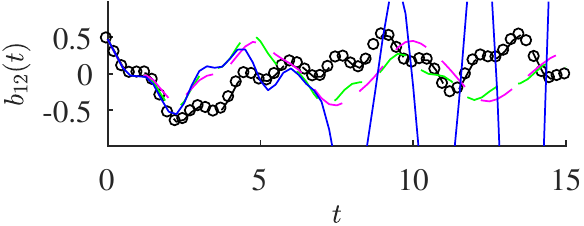}
	\includegraphics[width=0.45\linewidth]{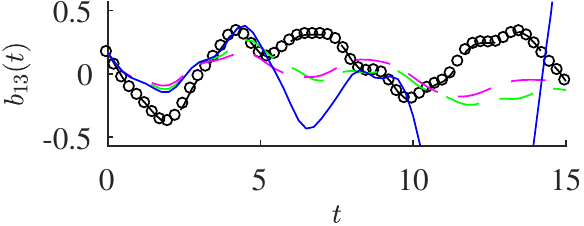}
	\includegraphics[width=0.45\linewidth]{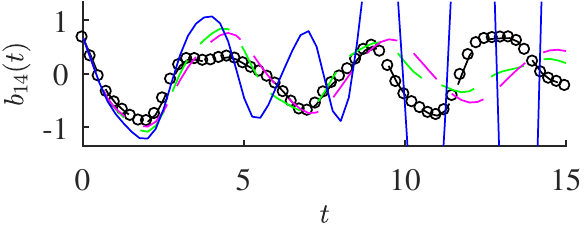}
	\includegraphics[width=0.45\linewidth]{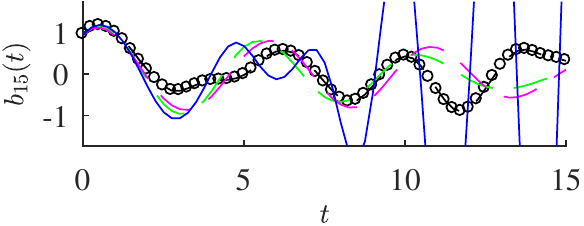}
	\includegraphics[width=0.45\linewidth]{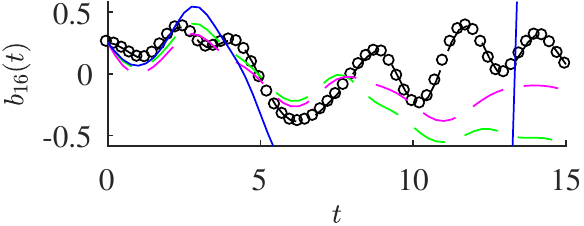}
	\caption{Reconstruction of the first ten \textit{chronos} for a wake flow at Reynolds number $Re=3900$ (HR LES), with $n=16$ POD modes and with a variance tensor expressed as a linear function of the \textit{chronos} and with modal characteristic times: (black dots), observed references; (blue line), standard POD-Galerkin; ( dashed magenta line), proposed stochastic representation; (dashed green line), modal eddy viscosity reduced-order model. The initial condition, at $t=0$, is identical for all methods.
	}
	\label{plot_3D_16m}
	\end{figure}
	
\begin{figure}
\centering
\includegraphics[width=\linewidth]{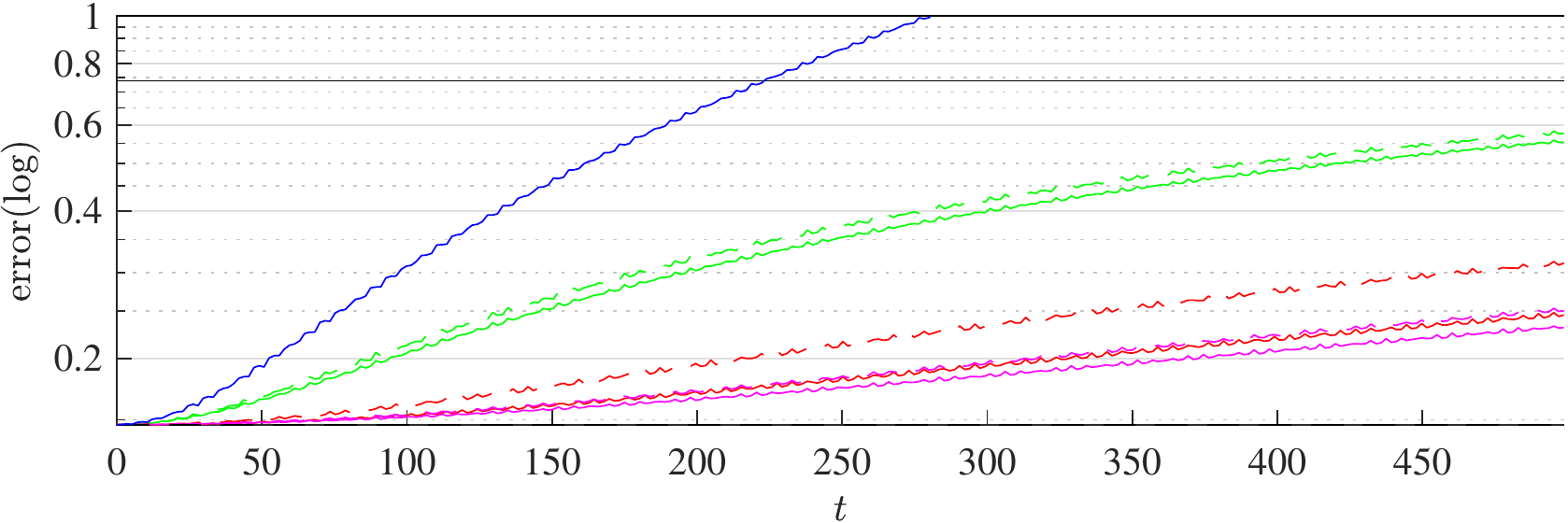}
\caption{Normalized reconstruction error for a wake flow at Reynolds number $Re=100$, with $n = 2$ POD modes:  (blue line), standard POD Galerkin (without eddy viscosity); (red/magenta lines), proposed model with a stationary/nonstationary variance tensor, solid and dashed line with single and modal characteristic time, respectively; (green lines), solid and dashed with eddy viscosity and modal eddy viscosity reduced-order model, respectively. The modal eddy viscosity coefficients are estimated through a least squares fit on the whole data sequence.  The dashed line indicates the error associated with the mode truncation : $\sum_{i=n+1}^N \lambda_i$. The black solid line is the error when we only consider the temporal mean velocity.}
\label{plot_logerr_2D100}
\end{figure}

\begin{figure}
	\centering
\includegraphics[width=\linewidth]{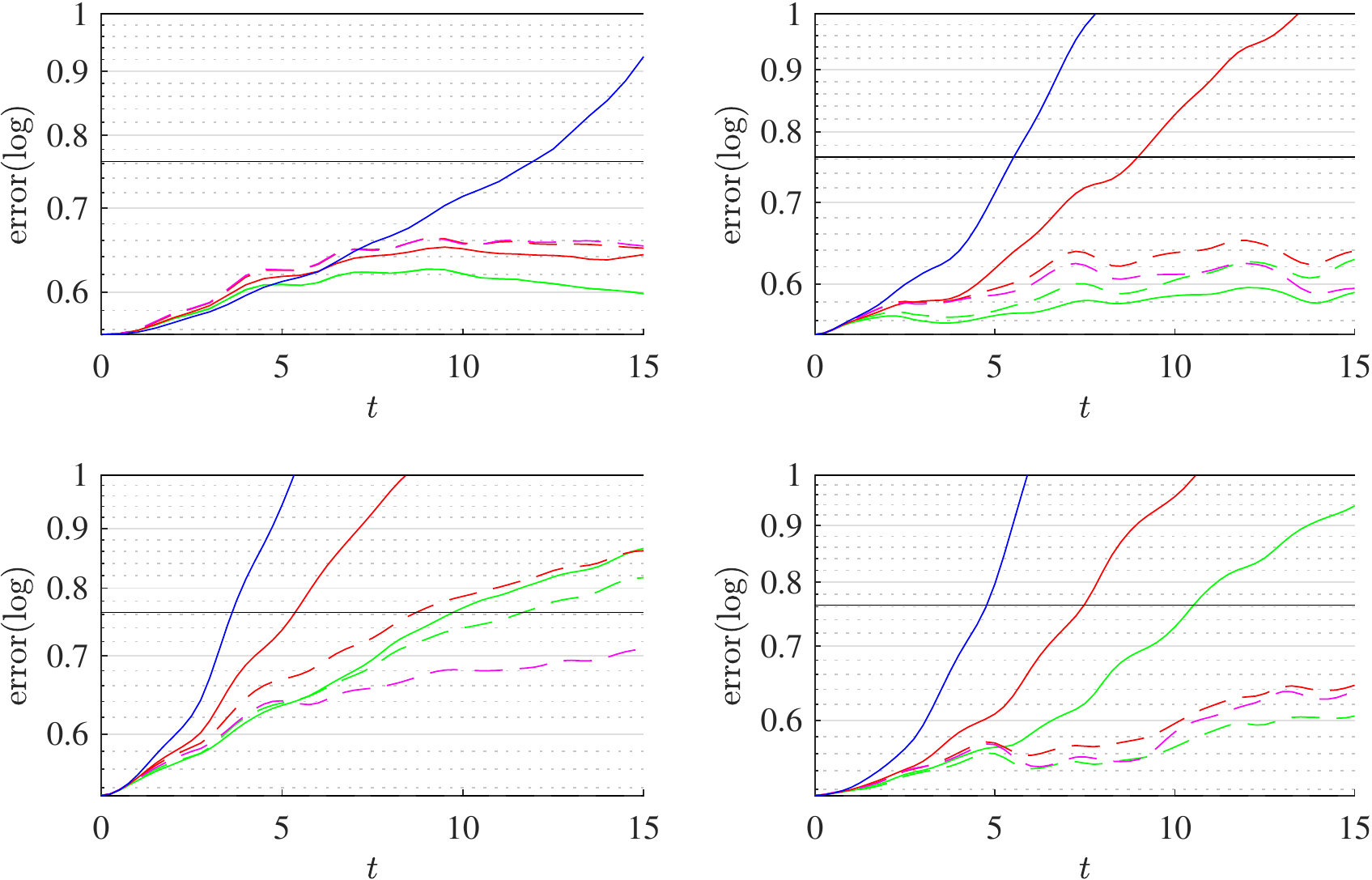}
	\caption{Normalized reconstruction error for a wake flow at Reynolds number $Re=3900$  (LES), with (from left to right and from top to bottom) $n$ = 2, 4,  8 and 16 POD modes: (blue line), standard POD Galerkin without eddy viscosity; (red/magenta lines), proposed model with a stationary/nonstationary variance tensor, solid and dashed line with single and modal characteristic time, respectively; (green lines), solid and dashed with eddy viscosity and modal eddy viscosity reduced-order model, respectively. The eddy viscosity coefficients are estimated through a least squares fit on the whole data sequence. The variance tensors are also estimated from a temporal mean on the whole sequence. The dashed line indicates the error associated with the mode truncation : $\sum_{i=n+1}^N \lambda_i$. The black solid line is the error when we only consider the temporal mean velocity.
}
	\label{plot_logerr_3DLES}
	\end{figure}

 Figures \ref{plot_2D_2m} and \ref{plot_3D_16m} show examples of the reconstruction of the \textit{chronos} for $n=2$ at Reynolds number $Re=100$ and $n=16$ at Reynolds number $Re=3900$, respectively, for the classical POD method (blue plot) and for the proposed modelling with respectively a stationary variance tensor (red line in figure \ref{plot_2D_2m}) and a modal nonstationary variance tensor defined on the subspace associated with the \textit{chronos} basis  (dashed magenta line in figure \ref{plot_3D_16m}) respectively. 
Each plot is sampled at the frequency $1/\Delta t$.
At Reynolds number $Re=100$, the first two modes carry most of the energy. The references $b^{ref}_i$ (black plots) and the \textit{chronos} obtained from an eddy viscosity model are superimposed for comparison purposes. 
The eddy viscosity is optimally fitted by a least squares estimation.
It can be observed that our stochastic model follows the references quite well whereas the deterministic model blows up.
Let us point out that here  our reduced models are completely parameter free unlike the eddy viscosity model.
Figures \ref{plot_logerr_2D100} and \ref{plot_logerr_3DLES} describe the error evolution along time. Approximating the square of the actual unresolved \textit{chronos} by the time average of their squares, we defined the error as follows:
\begin{eqnarray}
err (t)
&=& T \frac{\left  \|{\mbs v}^{ref} - {\mbs v} \right \|_{L^2(\Omega )} }{
\left  \|{\mbs u}^{ref}  \right \|_{L^2(\Omega \times [0,T])}} \nonumber ,\\
&=& 
T \frac{\left  \| \sum_{i=1}^n \left ( b_i^{ref} -  b_i \right ) {\mbs \phi}_i 
+  \sum_{i=n+1}^N b_i^{ref} {\mbs \phi}_i \right \|_{L^2(\Omega)} 
}{\left  \|\sum_{i=0}^N  b_i^{ref}  {\mbs \phi}_i  \right \|_{L^2(\Omega \times [0,T])}}
 ,\nonumber \\
& \approx &
 \left  (
 \frac{ \sum_{i=1}^n \left  ( b_i^{ref} -  b_i \right )^2
+ \sum_{i=n+1}^N  \lambda_i 
  }{
\left  \|{\mbs \phi}_0  \right \|^2_{L^2(\Omega)} + \sum_{i=1}^N \lambda_i 
}\right ) ^{1/2},
\label{approx_error}
\end{eqnarray}
which is greater  than the minimal error associated with the  modal truncation:
\begin{eqnarray}
\label{bound_error}
err (t) & \geqslant & \left  (
\frac{ \sum_{i=n+1}^N  \lambda_i 
  }{
\left  \|{\mbs \phi}_0  \right \|^2_{L^2(\Omega)} + \sum_{i=1}^N \lambda_i 
}\right ) ^{1/2}.
\end{eqnarray}

Equation (\ref{approx_error}) defines the criterion error plotted in figures \ref{plot_logerr_2D100} and \ref{plot_logerr_3DLES}, whereas (\ref{bound_error}) constitutes a lower bound of this error.
In figures \ref{plot_logerr_2D100} and \ref{plot_logerr_3DLES}, we successively displayed the error plots obtained for the standard POD Galerkin model without subgrid dissipative term, for our model with  stationary and nonstationary variance tensors, and finally for a deterministic modal eddy viscosity model. This subgrid model, proposed in \cite{Rempfer94} consists in modifying  the reduced-order system by adding a strong isotropic diffusive term (Laplacian) to stabilize the system. This eddy viscosity is said to be modal since different viscosity coefficients are attached to  each \textit{chronos}. Those coefficients are estimated by a least squares fitting on the  whole data sequence. Modal eddy viscosity in its least squares form resembles indeed a data assimilation strategy in which the best stationary isotropic dissipative forcing is estimated from the discrepancy between the model and the data. The same isotropic dissipation is imposed  on the whole fluid domain at every time step. As such this subgrid dissipation is much more difficult to interpret  in terms of local signatures of the small-scale coherent structures.

In figures \ref{plot_logerr_2D100} and \ref{plot_logerr_3DLES}, the dashed lines indicate the minimal error associated with the reduced subspace truncation error. The black solid line corresponds to the error level associated with  the temporal mean velocity -- i.e. setting all the \textit{chronos} to $0$. In this case:
\begin{eqnarray}
err_{| b=0} (t)
& \approx & 
 \left  (
\frac{ \sum_{i=1}^N  \lambda_i 
  }{
\left  \| \mbs \phi_0  \right \|^2_{L^2(\Omega)} + \sum_{i=1}^N \lambda_i 
}\right ) ^{1/2} .
\end{eqnarray}
This term does not constitute an upper bound of the error. However, it provides the error level reached by  the null model. In figures \ref{plot_logerr_2D100} and \ref{plot_logerr_3DLES}, fixing a log-scale for the $y$ axis, we readily observe the exponential divergence of the standard POD reduced-order model (in blue). 
	
We observe that even at Reynolds number $Re=100$ the modal eddy viscosity model does not capture accurately, on a long time period, the complex non-linear dynamics undergone  by the non-resolved modes (figure \ref{plot_logerr_2D100}). The eddy viscosity model over-damped the \textit{chronos} as shown in figure \ref{plot_2D_2m}.

For the  LES at Reynolds number $Re=3900$ (figure \ref{plot_logerr_3DLES}), we compared the eddy viscosity approaches (modal and constant) with stationary and nonstationary models of the variance tensors. In this case the variance tensor as well as the eddy viscosity coefficients have been estimated  with $1,460 \times \Delta t^{obs}/\Delta t$ vortex shedding cycles where $\Delta t^{obs}$ is the initial time step of the data (see table \ref{tab:cases})  whereas $\Delta t$ is the optimal subsampling time step given by the criterion \eqref{eq time step}.  The performances of the modal and single characteristic times  attached to the  variance tensor have been evaluated and compared. The error plots are shown in figure \ref{plot_logerr_3DLES}.
The introduction of different small-scale characteristic time steps associated with the different modes significantly improves the results that were obtained for a single common characteristic time. Both approaches are equivalent for short time period only. The introduction of modal characteristic times is clearly beneficial in the long run. The nonstationary representation performs  better than the stationary one especially for $n=8$ modes. Indeed, due to its non-stationarity the associated reduced system is even better than eddy viscosity models. Moreover, the piece of information brought by the nonstationary diffusion modes enables meaningful analysis of the small-scale contribution (see \S \ref{ssec:small-scale energy}).  Except for $n=8$ modes, the modal eddy viscosity approaches performs well. The constant eddy viscosity appears to only work
 when a small number of modes is involved. The (stationary or nonstationary) variance tensor models that are associated with modal characteristic time scales exhibit nearly the same stabilizing  skills  as the eddy viscosity models (again excepting the case $n=8$ modes). Both models lead to similar error levels. Nevertheless, it must be outlined that the two approaches are based on different assumptions. Eddy viscosity relies intrinsically on a homogeneous isotropic diffusion with no preferential direction of energy dissipation. The diffusion remains constant whatever the considered region: in the near or far wake regions, and even in the shear layers. However, as a fixed constant estimated through a mean squares  procedure, it provides the optimal amount of missing energy dissipation (with respect to a spatio-temporal mean of the squared norm) that is required to stabilize the reduced dynamical system.  Conversely, as shown in the previous section, the variance tensor and the associated diffusion modes provide a finer representation of the small-scales action in terms of energy dissipation but also in terms of energy redistribution. As for the simulation of the reduced system, both models often lead to comparable error levels. They bring stability to the system in a similar way, but the variance tensor models unveil important  clues on the small-scale flow structuration. 

\section{Conclusion}
We investigated the study of reduced-order modelling based on a stochastic representation of the small-scales proposed by \citet{memin2014fluid} and \citet{resseguier2016geo1}. This principle gives rise naturally to a drift correction generated by the inhomogeneity of the small-scale velocity  variance and  to an inhomogeneous diffusion term. The diffusion term is closely related to eddy viscosity assumption. Indeed, for an isotropic divergence-free random field, the stochastic representation  boils down to the classical eddy viscosity assumption. A POD Galerkin projection of the corresponding stochastic Navier-Stokes equations provides a modified reduced-order dynamical system that includes a linear term gathering the effects of the effective advection and of the diffusion exerted by the unresolved small-scale component. This function directly depends on the small-scale variance that must be specified to close the system.

 We proposed in this study a modelling based on the decomposition of this variance tensor on the {\em chronos} basis. The estimation has been performed on the residuals between the snapshots' measurements and their resolved reconstruction on the {\em topos} basis. The coefficients of this decomposition quoted as the {\em diffusion modes} constitute meaningful features for the interpretation of the small-scale statistical organization. They allow us to examine in details the principal directions of the large-scale energy dissipation and also to extract advective structures both generated by the small-scale velocity. 

The diffusion modes analysis has been applied to a circular cylinder wake flow. For this flow configurations in the laminar and in the subcritical regimes, the anisotropy of the diffusion modes determine regions of the flow that are key players.
 The largest magnitudes of the anisotropy zero-mode (stationary mode) occur both in the vicinity of the pivotal zone of the shear layers rolling into vortices and also where the drift correction is effective.

Finally, the diffusion modes were coupled with modal characteristic time scales to provide a subgrid model. For wake flows, stabilizing skills are comparable to those obtained with optimally identified isotropic eddy viscosity models. Such a stochastic approach consisting of a rigorously derived subgrid term may easily be
applied to other turbulent flow configurations, for instance boundary layer flows that are known to develop complex multi-scale mechanisms. Moreover, it will be interesting to see whether our proposed stochastic POD models could be used to design novel physics-based subgrid scale models for LES approaches.

In order to obtain better \textit{chronos} reconstruction results than that of an eddy viscosity model, random forcings can be included -- as in equation \eqref{SST} -- in the stochastic Navier-Stokes representation \eqref{NSvf}. This additional complexity maintains the variability of stable temporal modes. This full stochastic extension will be the subject of a next study.
Other extensions of this methodology include online quantification of model errors for ensemble data assimilation procedures. The authors are already pursuing these works in the context of reduced order models but also in geophysical fluid dynamics \citep{resseguier2016geo1,resseguier2016geo2,Yang_Memin_2017}.


\section*{Acknowledgements}
The authors acknowledge the support of the SEACS project funded by the  Brittany clusters of excellence (Cominslab, Lebesgue, and Mer).
The authors thank Pranav Chandramouli for providing the data case at Reynolds $\Rey=3900$.

\bibliographystyle{jfm}
\bibliography{biblio}


\end{document}